\documentclass[aps,pre,superscriptaddress,longbibliography,twocolumn]{revtex4-2}
\usepackage{amsmath}
\usepackage{amsfonts}
\usepackage{amssymb}
\usepackage{lmodern,dsfont}
\usepackage{graphicx}
\usepackage[usenames,dvipsnames]{xcolor}
\usepackage{bm}
\usepackage[english=american]{csquotes}
\usepackage{xr}
\usepackage{tikz}

\newcommand{\kb}{k_\text{B}}
\newcommand{\Av}[1]{\left\langle #1 \right\rangle}
\newcommand{\av}[1]{\langle #1 \rangle}
\newcommand{\n}{\nonumber}
\newcommand{\nn}{\nonumber \\}

\newcommand{\grad}{\bm{\nabla}}
\newcommand{\Red}[1]{{\color{Red} #1}}
\renewcommand{\eqref}[1]{Eq.~(\ref{#1})}
\newcommand{\bmc}[1]{\bm{\mathcal{#1}}}

\newcommand{\estimates}{\mathrel{\hat=}}

\begin{document}

\title{Entropy production and non-Gaussianity of fast processes at weak damping}

\author{Mario A. Ciampini}
\affiliation{Vienna Center for Quantum Science and Technology, Faculty of Physics, University of Vienna, Vienna A-1090, Austria}
\author{Jakob Rieser}
\affiliation{Vienna Center for Quantum Science and Technology, Faculty of Physics, University of Vienna, Vienna A-1090, Austria}
\author{Nikolai Kiesel}
\affiliation{Vienna Center for Quantum Science and Technology, Faculty of Physics, University of Vienna, Vienna A-1090, Austria}
\author{Andreas Dechant}
\affiliation{Department of Physics \#1, Graduate School of Science, Kyoto University, Kyoto 606-8502, Japan}

\begin{abstract}
We present a method of estimating the rate of entropy production in underdamped dynamics by decomposing it into contributions originating in different non-equilibrium effects.
Specifically, a non-zero average velocity, a non-thermal width of the velocity distribution, correlations between position and velocity and non-Gaussian velocity statistics represent different ways in which the system can be out of equilibrium and each give rise to a positive contribution to the overall entropy production rate.
We demonstrate that each contribution can be separately estimated from experimental trajectory data of levitated nano-particles subject to non-linear forces.
We find that the majority of the entropy production rate can be attributed to the first three contributions which can be estimated from the first and second moments of the position and velocity and therefore result in a useful \enquote{Gaussian} estimate for the entropy production rate.
\end{abstract}

\date{\today}

\maketitle

\section{Introduction} \label{sec-intro}
Entropy production is an unavoidable aspect of all physical systems out of thermodynamic equilibrium.
On a fundamental level, it characterizes the breaking of time-reversal symmetry in the dynamics of the system as a consequence of its interactions with its environment.
From a more practical point of view, it also manifests in the form of dissipation, leading to unrecoverable loss of energy into the environment and preventing us from realizing processes at perfect efficiency.
Both statements are quantitative; the amount of entropy production both characterizes the degree of irreversibility and, together with the temperature of the environment, determines the amount of dissipated energy.
Thus, for both fundamental and practical reasons, we are interested in a precise estimation of the entropy production of a physical system and the estimation of the overall entropy production has been a central topic of interest in recent research.
In particular, sparked by key discoveries that employ the statistics of measured currents to obtain lower bounds on the amount of entropy production \cite{Barato2015,Gingrich2016,Pietzonka2018,Dechant2018,Horowitz2020,Dechant2021a}, methods that infer the entropy production rate by optimizing over currents have been developed \cite{Manikandan2020,Otsubo2020}.
Other observations that can be used obtain lower estimates on the entropy production include the statistics of transition events in discrete systems \cite{Meer2022,Harunari2022}.

While thermodynamic equilibrium places a strong constraint on a physical system, out-of-equilibrium systems are much more diverse.
For a single given equilibrium system, there are generally many physically distinct ways of driving it out of equilibrium.
Since all of them make the dynamics irreversible and thus result in entropy production, a natural question is whether we can identify distinct contributions to the latter that correspond to different mechanisms of driving.
One of the most well-known examples is the decomposition of entropy production into excess and housekeeping contributions \cite{Hatano2001,Maes2014}, which separates the dissipation due to time-dependent and non-conservative driving forces. This approach been generalized based on concepts rooted in geometry, optimal transport and information theory \cite{Nakazato2021,Dechant2022,Yoshimura2023}.
More recently, decompositions separating the contributions of different spatial or temporal modes of the dynamics to entropy production have also been obtained \cite{Nagayama2023,Sekizawa2024}.

A serious limitation of both estimation and decomposition of entropy production is that most methods are only applicable to systems at strong damping, so-called overdamped systems.
The underlying reason is that, in this class of systems, any change in the configuration of the system results in dissipation, so there is a direct correspondence between the observed dynamics and entropy production.
By contrast, at weak damping, we also have to consider changes in the phase-space configuration that correspond to the reversible, Hamiltonian dynamics of the system, which do not induce any dissipation.
The interplay between reversible and irreversible dynamics makes the description of weakly damped systems much more challenging.
In particular, many of the inequalities derived for overdamped systems do not apply \cite{Pietzonka2022} and, while a decomposition into excess and housekeeping entropy can be formally defined, its physical interpretation is less clear \cite{Spinney2012,Spinney2012a,Lee2013,Sasa2014}.

\begin{figure*}
    \includegraphics[width=.9\textwidth]{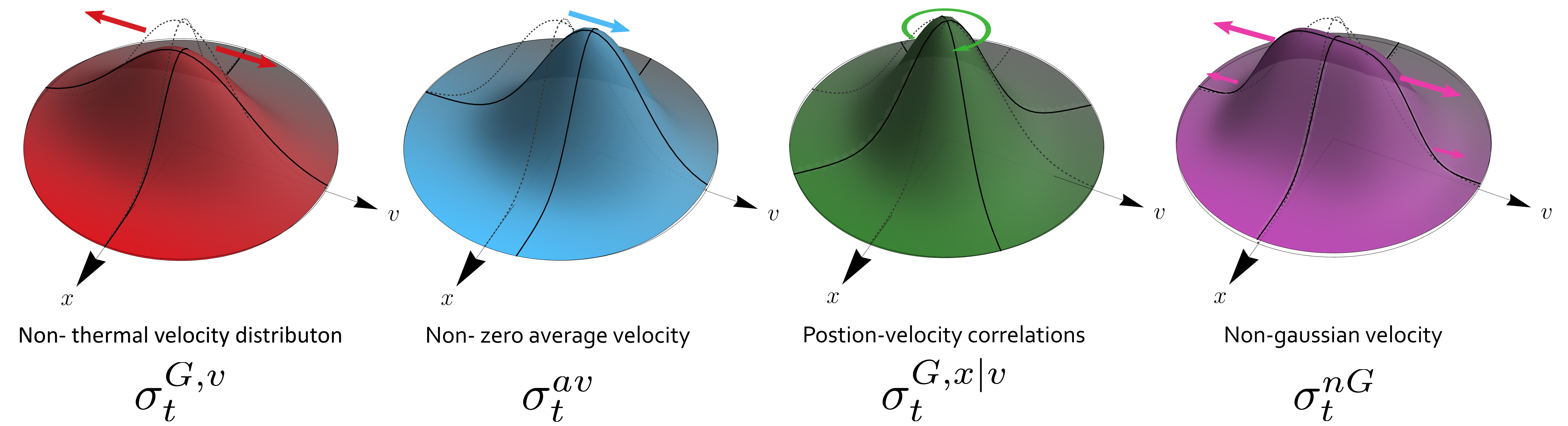}
    \caption{Illustration of different types of non-equilibrium phase-space densities for and underdamped system. In contrast to the equilibrium phase-space density (blue, center), a non-equilibrium system non-zero average velocity (density displaced from $v = 0$, top right), a width that is different from the thermal velocity (density stretched or compressed in $v$-direction, top left), correlations between position and velocity (density rotated in the $x$-$v$ plane, bottom left) or a non-Gaussian velocity density (more weight in the tails or center of the density, bottom right). Our main result is that each of these features gives rise to a separate, positive contribution to the entropy production rate of the system, which sum to the total entropy production rate.}
    \label{fig-illustration}
\end{figure*}

In this work, we focus on the experimental system of optically levitated nano-particles, which operates at weak damping.
Small particles are trapped at an intensity maximum of a standing light wave by dipole forces. 
The surrounding air acts as a thermal environment, with which the particles interact via collisions with the air molecules.
In addition, electrodes allow fast control over the electrostatic force acting on charged particles.
This setup, on the one hand, allows monitoring the particle position with sufficient temporal and spatial resolution to access its phase-space dynamics.
On the other hand, since the parameters of the light field can be adjusted with high speed and precision, the setup allows implementing fast far-from-equilibrium processes.
Crucially, these experiments are performed at low environmental pressure, and thus in the weak damping regime: the frequency of oscillations of the particle in the trap is orders of magnitude larger than the thermal damping rate.

Our goal is to use measured trajectory data to estimate the entropy production of weakly damped systems such as levitated nano-particles under the influence of time-dependent and generally non-linear forces driving it out of equilibrium.
Our main result is that we can decompose the entropy production into four distinct, positive contributions, which reflect different non-equilibrium properties of the phase-space statistics of the particles, see Fig.~\ref{fig-illustration}. 
Specifically, we can attribute the contributions to a non-zero average velocity, non-thermal velocity covariance, correlations between position and velocity and a non-Gaussian velocity distribution, respectively.

Crucially, all four contributions can be obtained separately from experimental trajectory data. 
Besides estimating the overall entropy production of weakly-damped particles, this provides insight into the reason for the non-equilibrium conditions, allowing us to distinguish between different ways that the system is being driven out of equilibrium.
Three of the four contributions can be obtained from the first- and second-order moments of the position and velocity of the particle, which are straightforward to estimate from the data and provide a quantitatively useful \enquote{Gaussian} lower estimate on the entropy production rate.
The fourth, non-Gaussian contribution is evaluated using a variational representation of Fisher information.

The remainder of the article is structured as follows:
In Section \ref{sec-langevin} we review the description of weakly-damped motion using the underdamped Langevin and Kramers equation and its thermodynamic interpretation in terms of heat and entropy.
In Section \ref{sec-estimation-bounds} we derive our main result, the decomposition of the entropy production rate and the variational estimation of Fisher information and show how this can be used to infer bounds on the entropy production rate.
We briefly introduce the details of the experiment in Section \ref{sec-experiment}, before applying the results to the experimentally obtained trajectory data in Section \ref{sec-data-analysis}.

\section{Underdamped Langevin dynamics} \label{sec-langevin}
We consider a $d$-dimensional system of (possibly interacting) particles with coordinates $(x_1,\ldots,x_d) = \bm{x}$ and velocities $(v_1,\ldots,v_d) = \bm{v}$ that are in contact with a viscous equilibrium environment characterized by the friction coefficient $\gamma$ and the temperature $T$.
The motion of the particles is described by the underdamped Langevin equation \cite{Coffey2017}
\begin{align}
\dot{\bm{x}}(t) = \bm{v}(t), \quad m \dot{\bm{v}}(t) = -\gamma \bm{v} + \bm{f}_t(\bm{x}(t)) + \sqrt{2 \gamma \kb T} \bm{\xi}(t) \label{langevin} .
\end{align}
For simplicity, we assume that the particles have the same mass $m$ and friction coefficient $\gamma$; an extension to more general situations is straightforward.
From here on out, we also set $\kb = 1$ to simplify the notation.
The force $\bm{f}_t(\bm{x})$ describes both external forces and interactions between the particles, it may depend explicitly on time via an external control.
$\bm{\xi}(t)$ is a vector of mutually independent, Gaussian white noises that describes the randomization of the particles' velocities due to collisions with the molecules of the environment.
For now, we assume that the friction force and the noise satisfy the fluctuation-dissipation relation; we will treat a more general setup later when discussing the experiment.
While \eqref{langevin} directly describes the noisy trajectories of the system, we can equivalently describe the statistics of these trajectories using the probability density $p_t(\bm{x},\bm{v})$, which measures the probability of observing a configuration $(\bm{x},\bm{v})$ at time $t$.
The latter evolves according to the Kramers-Fokker-Planck equation \cite{Risken1986,Coffey2017}
\begin{align}
\partial_t p_t(\bm{x},\bm{v}) &= - \grad_x \cdot \bm{j}_t^{x,\text{rev}}(\bm{x},\bm{v}) \\
& - \grad_v \cdot \big( \bm{j}_t^{v,\text{rev}}(\bm{x},\bm{v}) + \bm{j}_t^{v,\text{irr}}(\bm{x},\bm{v}) \big) \quad \text{with} \nn
\bm{j}_t^{x,\text{rev}}(\bm{x},\bm{v}) &= \bm{v} p_t(\bm{x},\bm{v}) \nn
\bm{j}_t^{v,\text{rev}}(\bm{x},\bm{v}) &= \frac{1}{m} \bm{f}_t(\bm{x}) p_t(\bm{x},\bm{v}) \nn
\bm{j}_t^{v,\text{irr}}(\bm{x},\bm{v}) &= -\frac{\gamma}{m} \bigg( \bm{v} + \frac{T}{m} \grad_v \bigg) p_t(\bm{x},\bm{v}) \n .
\end{align}
Here $\bm{j}_t^{x/v}$ denotes the probability current in the $\bm{x}$-, respectively $\bm{v}-$, direction.
The reversible probability currents (indicated as rev) correspond to the deterministic motion of the system, whereas the irreversible probability currents (indicated as irr) describe the interaction with the environment.
Crucially, the irreversible currents also determine the rate of entropy production, \cite{Spinney2012,Spinney2012a,Lee2013}
\begin{align}
\sigma_t = \frac{m^2}{\gamma T} \Av{ \frac{\Vert \bm{j}_t^{v,\text{irr}} \Vert^2}{p_t^2}}_t,
\end{align}
where $\Av{\ldots}_t$ denotes an average with respect to $p_t(\bm{x},\bm{v})$.
$\sigma_t$ measures the rate at which the total entropy of the system and its environment increases; it is positive whenever the irreversible currents are non-zero.
We can decompose the total entropy production rate into the rate of change of (Gibbs-Shannon) entropy of the system and the change in entropy of the environment due to heat flows between the system and the latter \cite{Sekimoto2010},
\begin{align}
\sigma_t &= \sigma_t^\text{sys} + \sigma_t^\text{env} \quad \text{with} \label{entropy-system-environment} \\
\sigma_t^\text{sys} &= d_t S_t = - d_t \Av{\ln p_t}_t \nn
\sigma_t^\text{env} &= \frac{\gamma}{m} \bigg( \frac{m \Av{\Vert \bm{v} \Vert^2}}{T} - d \bigg) \n .
\end{align}
We define the (positive definite) covariance matrix of the velocities as
\begin{align}
\big(\bm{\Xi}^v_t \big)_{kl} = \text{Cov}_t(v_k,v_l) = \Av{v_k v_l}_t - \Av{v_k}_t \Av{v_l}_t .
\end{align}
With this, we can write the rate of entropy change in the environment as
\begin{align}
\sigma_t^\text{env} = \frac{\gamma}{T} \big\Vert \av{\bm{v}}_t \big\Vert^2 + \frac{\gamma}{m} \bigg( \frac{m}{T} \text{tr}\big(\bm{\Xi}_t^v \big) - d \bigg) \label{entropy-environment-covariance} ,
\end{align}
where tr denotes the trace of a matrix.
Rewriting the rate of change of the system's entropy, we find a similar expression
\begin{align}
\sigma_t^\text{sys} = \frac{\gamma}{m} \bigg( \frac{T}{m} \text{tr}\big(\bm{F}_t^v \big) - d \bigg) \label{entropy-shannon-fisher},
\end{align}
where $\bm{F}_t^v$ is the so-called Fisher information matrix of the probability density with respect to velocity.
\begin{align}
\big(\bm{F}_t^v \big)_{kl} = \Av{\partial_{v_k} \ln p_t \partial_{v_l} \ln p_t}_t = - \Av{\partial_{v_k} \partial_{v_l} \ln p_t}_t .
\end{align}
Intuitively, the positive definite matrix $\bm{F}_t^v$ measures how sensitive the probability density is to changes in the velocity; it is large if $p_t(\bm{x},\bm{v})$ has sharp features as a function of $\bm{v}$, e.~g.~for a very narrow distribution.
Thus, while the covariance matrix $\bm{\Xi}_t^v$ determines the change in entropy of the environment, the Fisher information $\bm{F}_t^v$ determines the change in the system's entropy.
Generally, if the velocity distribution is narrow (compared to the equilibrium width $T/m$) then $\bm{\Xi}_t^v$ is small, while $\bm{F}_t^v$ is large.
Therefore, the system absorbs heat from the environment, $\sigma_t^\text{env} < 0$, which causes the system to thermalize and the width of the distribution to increase, $\sigma_t^\text{sys} > 0$.
Conversely, for a wide distribution, $\bm{\Xi}_t^v$ is large, whereas $\bm{F}_t^v$ is small, and the system dissipates heat into the environment, causing the distribution to narrow, $\sigma_t^\text{env} > 0$ and $\sigma_t^\text{sys} < 0$.
During both processes, the second law implies that $\sigma_t^\text{env} + \sigma_t^\text{sys} \geq 0$ at any point in time, so that the increase in one of the two quantities always dominates the decrease in the other.

One interesting consequence of \eqref{entropy-shannon-fisher} is that, if the system is in a steady state, where we have $\sigma_t^\text{sys} = 0$, the Fisher information matrix with respect to velocity satisfies the condition
\begin{align}
\text{tr}\big(\bm{F}_\text{st}^v \big) = \frac{d m}{T} \label{fisher-steady-temperature} .
\end{align}
This places a constraint on what probability distributions can be observed as steady states of an underdamped Langevin system.
This is in contrast to the overdamped case, where any distribution can in principle be realized by an appropriate choice of the force.
A practical application of this relation is that it allows to determine the environmental temperature from a measurement of the velocity statistics of a non-equilibrium steady state.
In equilibrium, the temperature can be obtained by measuring the average kinetic energy (second moment of the velocity) of the particles.
Out of equilibrium however, the latter is different from the environmental temperature and only satisfies the inequality
\begin{align}
\Av{E_\text{kin}}_\text{st} = \frac{m}{2} \text{tr}\big(\bm{\Xi}_\text{st}^v \big) \geq d T,
\end{align}
which follows from the positivity of the entropy production rate \eqref{entropy-system-environment} and cannot be used to directly infer the temperature.
By contrast, \eqref{fisher-steady-temperature} implies that, provided that we can observe the non-thermal velocity statistics precisely enough to compute their Fisher information, the latter can be used to infer the environmental temperature.

\section{Estimation and bounds for entropy production rate} \label{sec-estimation-bounds}

\subsection{Estimating Fisher information and entropy rate from data} \label{sec-variational}
From the point of view of estimating the entropy production rate from measurement data, an estimation of the entropy change in the environment is straightfoward, since it only requires estimating the covariance matrix of the velocity, provided that the mass, friction coefficient, and temperature are known.
However, the entropy change of the system is more subtle, since it explicitly depends on the probability density $p_t(\bm{x},\bm{v})$ which is difficult to estimate accurately unless we have an exceedingly large amount of data.
This is complicated even more by the fact that we have to estimate $p_t(\bm{x},\bm{v})$ accurately enough to take its derivative, either with respect to time or with respect to the velocity,
\begin{align}
\sigma_t^\text{sys} = - \Av{\partial_t \ln p_t}_t = \frac{\gamma}{m} \bigg( \frac{T}{m} \Av{\Vert \grad_v \ln p_t \Vert^2}_t - d \bigg),
\end{align}
where we used the expressions \eqref{entropy-system-environment} and \eqref{entropy-shannon-fisher}, respectively.
However, the Fisher information matrix entering the latter expression can be more conveniently evaluated using a variational representation.
Let us consider some arbitrary smooth function $\phi(\bm{x},\bm{v})$ and write,
\begin{align}
\int d\bm{x} &\int d\bm{v} \ \grad_v \phi(\bm{x},\bm{v}) \cdot \grad_v \ln p_t(\bm{x},\bm{v}) p_t(\bm{x},\bm{v}) \\
& = - \int d\bm{x} \int d\bm{v} \ \grad_v^2 \phi(\bm{x},\bm{v}) p_t(\bm{x},\bm{v}) = \Av{ \grad_v^2 \phi}_t , \n
\end{align}
where we integrated by parts with respect to $\bm{v}$.
We also have from the Cauchy-Schwarz inequality
\begin{align}
\Big(\Av{\grad_v \phi \cdot \grad_v \ln p_t}_t \Big)^2 \leq \Av{\Vert \grad_v \phi \Vert^2}_t \Av{\Vert \grad_v \ln p_t \Vert^2}_t .
\end{align}
Combining the two relations, we have
\begin{align}
\Av{\Vert \grad_v \ln p_t \Vert^2}_t = \text{tr}\big(\bm{F}_t^v \big) \geq \frac{\big(\Av{ \grad_v^2 \phi}_t \big)^2}{\Av{\Vert \grad_v \phi \Vert^2}_t} .
\end{align}
Equality is attained by the choice $\phi(\bm{x},\bm{v}) = \ln p_t(\bm{x},\bm{v})$ and thus,
\begin{align}
\text{tr}\big(\bm{F}_t^v \big) = \sup_\phi \Bigg[ \frac{\big(\Av{ \grad_v^2 \phi}_t \big)^2}{\Av{\Vert \grad_v \phi \Vert^2}_t} \Bigg] \label{fisher-variational} ,
\end{align}
where $\sup$ denotes the supremum.
Crucially, the right-hand side only involves taking averages over derivatives of the known function $\phi(\bm{x},\bm{v})$; in other words, we have shifted the task of computing derivatives of the probability density, to computing derivatives of known functions and maximizing over functions.
This directly yields a variational expression for the rate of entropy change
\begin{align}
\sigma_t^\text{sys} = \frac{\gamma}{m} \Bigg( \frac{T}{m} \sup_\phi \Bigg[ \frac{\big(\Av{ \grad_v^2 \phi}_t \big)^2}{\Av{\Vert \grad_v \phi \Vert^2}_t} \Bigg] - d \Bigg) .
\end{align}
An expression that is equivalent to \eqref{fisher-variational} but more immediately applicable is
\begin{align}
\text{tr}\big(\bm{F}_t^v \big) = \sup_\phi \Big[ 2 \Av{\grad_v^2 \phi}_t - \Av{\Vert \grad_v \phi \Vert^2}_t \Big] \label{fisher-variational-2} .
\end{align}
To see the equivalence, we note that \eqref{fisher-variational} is obtained by rescaling the variational function by a constant, $\phi(\bm{x},\bm{v}) \rightarrow \alpha \phi(\bm{x},\bm{v})$, and then maximizing with respect to $\alpha$.
In practice, we prescribe a set of basis functions $\psi_q(\bm{x},\bm{v})$, $q \in \lbrace 1, \ldots, Q \rbrace$ and write
\begin{align}
\phi(\bm{x},\bm{v}) = \sum_q c_q \psi_q(\bm{x},\bm{v}), \quad c_q \in \mathbb{R} .
\end{align}
Plugging this into \eqref{fisher-variational-2} and taking the derivative with respect to the coefficients $c_q$, we see that finding the optimal coefficients is equivalent to solving the set of linear equations
\begin{align}
\sum_r \Av{\grad_v \psi_q \cdot \grad_v \psi_r}_t c_r = \Av{\grad_v^2 \psi_q}_t \label{c-equations} .
\end{align}
Defining
\begin{align}
\big( \bm{D} \big)_{qr} = \Av{\grad_v \psi_q \cdot \grad_v \psi_r}_t, \quad \big( \bm{b} \big)_q = \Av{\grad_v^2 \psi_q}_t,
\end{align}
we then obtain
\begin{align}
\text{tr}\big(\bm{F}_t^v \big) \geq \bm{b} \cdot \bm{D}^{-1} \bm{b} \label{fisher-estimate} ,
\end{align}
where equality holds if the set of basis functions is complete.
Note that the left-hand side is equal to the trace of the Fisher information, so that we can write for the rate of change of the system's entropy
\begin{align}
\sigma_t^\text{sys} \geq \frac{m}{\gamma} \bigg( \frac{T}{m} \bm{b} \cdot \bm{D}^{-1} \bm{b} - d \bigg) \label{entropy-system-estimate} ,
\end{align}
where equality is achieved for a complete set of basis functions.
Thus, in order to estimate the change in the system's entropy, we first choose an appropriate set of basis functions, computing their first and second derivatives with respect to $\bm{v}$, and then average the latter with respect to the data set, yielding $\bm{D}$ and $\bm{b}$.
These can then be used to calculate the estimate \eqref{entropy-system-estimate}.
For later use, we note that we can also derive the more general relation for an arbitrary positive definite matrix $\bm{B} \in \mathbb{R}^{d\times d}$,
\begin{align}
\text{tr} \big( \bm{B} \bm{F}_t^v \big) = \sup_{\phi} \Bigg[ \frac{\text{tr} \big( \bm{B} \Av{\grad_v \grad_v^\text{T} \phi}_t \big)^2}{\text{tr} \big( \bm{B} \Av{\grad_v \phi \grad_v^\text{T} \phi}_t \big)} \Bigg] \label{fisher-variational-general},
\end{align}
which reduces to \eqref{fisher-variational} if $\bm{B}$ is the identity matrix.
Here, the superscript T denotes the transpose of a vector or matrix.
Again introducing a set of basis functions, we obtain,
\begin{align}
\text{tr} \big( \bm{B} \bm{F}_t^v \big) &\geq \bm{b}_B \cdot \bm{D}_B^{-1} \bm{b}_B \quad \text{with}  \\
\big( \bm{D}_B \big)_{qr} &= \text{tr}\big( \bm{B}  \Av{ \grad_v \psi_q \grad_v^\text{T} \psi_r}_t \big), \nn
\big( \bm{b}_B \big)_q &= \text{tr} \big(\bm{B} \Av{\grad_v \grad_v^\text{T} \psi_q}_t \big) , \n
\end{align}
which likewise reduces to the previous result if $\bm{B}$ is the identity.

The quality of the above estimate on the Fisher information depends on the choice of the basis functions.
On the one hand, we want the set of basis functions to be as complete as possible to assure that we can accurately approximate the optimal variational function $\phi(\bm{x},\bm{v}) = \ln p_t(\bm{x},\bm{v})$.
On the other hand, we also need to choose the basis functions such their averages with respect to the data set accurately reflect their true statistical averages; for example, we want to avoid functions whose derivatives are very large for rare values of $(\bm{x},\bm{v})$.
Moreover, choosing a too large set of basis functions may make the matrix $\bm{D}$ ill-conditioned, preventing an accurate numerical solution of \eqref{c-equations}.
We will investigate these issues in detail when applying our formalism to experimental data.

Finally, we remark that we can also obtain a direct variational expression for the entropy of the system,
\begin{align}
S_t = - \sup_{\rho} \Big[ \Av{\ln \rho}_t \Big],
\end{align}
where the maximum is taken over all probability densities $\rho(\bm{x},\bm{v})$.
This formula follows by noting that the Kullback-Leibler divergence between $p_t(\bm{x},\bm{v})$ and $\rho(\bm{x},\bm{v})$ is positive
\begin{align}
\Av{\ln \bigg( \frac{p_t}{\rho} \bigg)}_t = D_\text{KL}(p_t \Vert \rho) \geq 0,
\end{align}
and vanishes for $p_t(\bm{x},\bm{v}) = \rho(\bm{x},\bm{v})$.
From this, we have
\begin{align}
S_t \leq - \Av{\ln \rho}_t,
\end{align}
with equality for $\rho(\bm{x},\bm{v}) = p_t(\bm{x},\bm{v})$.
However, due to the presence of the logarithm, this variational expression is generally hard to evaluate; in particular, if we write
\begin{align}
\rho(\bm{x},\bm{v}) = \sum_q c_q \rho_q(\bm{x},\bm{v}),
\end{align}
where $\rho_q(\bm{x},\bm{v})$ is a basis set of probability densities and $c_q \geq 0$ with $\sum_q c_q = 1$ are coefficients, then we cannot obtain an explicit expression for $c_q$.
Moreover, this procedure only yields the instantaneous entropy $S_t$; in order to obtain its rate of change, we still have to differentiate with respect to time.

\subsection{Gaussian and non-Gaussian entropy production} \label{sec-decomposition}
In the above, we showed how the variational expression \eqref{fisher-variational} can be used to estimate the Fisher information and thus entropy rate from data.
As we show in the following, it also implies different decompositions of the entropy rates.
We start by noting that, choosing $\phi(\bm{x},\bm{v})$ in \eqref{fisher-variational} as
\begin{align}
\phi(\bm{x},\bm{v}) = \frac{1}{2} \big(\bm{v} - \Av{\bm{v}}_t \big)^\text{T} \big(\bm{\Xi}_t^v \big)^{-1} \big(\bm{v} - \Av{\bm{v}}_t \big),
\end{align}
we obtain
\begin{align}
\text{tr}\big(\bm{F}_t^v \big) \geq \text{tr}\big((\bm{\Xi}_t^v)^{-1} \big) \label{fisher-covariance} .
\end{align}
We can obtain a more refined version of this inequality as follows: 
First, we define the covariance and Fisher information matrices with respect to both position and velocity,
\begin{align}
\bm{\Xi}_t = \begin{pmatrix} \bm{\Xi}_t^x & \bm{\Xi}_t^{xv} \\ \bm{\Xi}_t^{xv,\text{T}} & \bm{\Xi}_t^v  \end{pmatrix}, \quad \bm{F}_t = \begin{pmatrix} \bm{F}_t^x & \bm{F}_t^{xv} \\ \bm{F}_t^{xv,\text{T}} & \bm{F}_t^v  \end{pmatrix} \label{fisher-covariance-xv} ,
\end{align}
both of which are positive definite.
In \eqref{fisher-variational-general}, we choose
\begin{align}
\phi(\bm{x},\bm{v}) = \frac{1}{2} \big( \bm{z} - \Av{\bm{z}}_t \big)^\text{T} \bm{\Xi}_t^{-1} \big( \bm{z} - \Av{\bm{z}}_t \big),
\end{align}
where we defined $\bm{z} = (\bm{x},\bm{v})$.
We obtain
\begin{align}
\text{tr} \big( \bm{B} \bm{F}_t \big) \geq \text{tr}\big( \bm{B} \bm{\Xi}_t^{-1} \big).
\end{align}
Since $\bm{B}$ is arbitrary, we can choose $\bm{B} = \bm{a} \bm{a}^\text{T}$ with some constant vector $\bm{a}$, which gives us
\begin{align}
\bm{a}^\text{T} \bm{F}_t \bm{a} \geq \bm{a}^\text{T} \bm{\Xi}_t^{-1} \bm{a} 
\end{align}
for any vector $\bm{a}$.
This is equivalent to the matrix inequality
\begin{align}
\bm{F}_t \geq \bm{\Xi}_t^{-1},
\end{align}
which is interpreted as $\bm{F}_t - \bm{\Xi}_t^{-1}$ being a positive semidefinite matrix.
Since an inequality between two matrices also implies an inequality on any of their principal submatrices (diagonal blocks), we have
\begin{align}
\bm{F}_t^v \geq \big(\bm{\Xi}_t^{-1} \big)^v .
\end{align}
Further, for a positive definite matrix, any diagonal block of the inverse is larger than the inverse of the diagonal block, so we obtain the chain of inequalities
\begin{align}
\bm{F}_t^v \geq \big(\bm{\Xi}_t^{-1} \big)^v \geq \big( \bm{\Xi}_t^v \big)^{-1} \label{fisher-covariance-inequality-xv}.
\end{align}
Note that the first inequality is an equality if the joint distribution of $\bm{x}$ and $\bm{v}$ is Gaussian; in this case we have $\bm{F}_t = \bm{\Xi}_t^{-1}$.
The second inequality is an equality if there are no correlations between $\bm{x}$ and $\bm{v}$, $\bm{\Xi}_t^{xv} = 0$.

We now rewrite \eqref{entropy-shannon-fisher}, as
\begin{align}
\sigma_t^\text{sys} &= \frac{\gamma}{m} \text{tr}\bigg(\frac{T}{m}  \big(\bm{\Xi}_t^{-1} \big)^{v} - \bm{I} \bigg) + \sigma_t^\text{nG} \quad \text{with} \label{shannon-decomposition} \\
\sigma_t^\text{nG} &= \frac{\gamma T}{m^2} \text{tr}\Big( \bm{F}_t^v - \big(\bm{\Xi}_t^{-1} \big)^{v} \Big) \n .
\end{align}
This separates the rate of entropy change into two terms: The first one only involves the covariance matrix and is therefore independent of whether the probability distribution is Gaussian or not; this term can be positive or negative.
The second term $\sigma_t^\text{nG}$, on the other hand, is positive because of \eqref{fisher-covariance-inequality-xv} and vanishes only if the joint distribution of $\bm{x}$ and $\bm{v}$ is Gaussian; it therefore directly expresses the influence of non-Gaussian statistics.
For the total entropy production rate, we have
\begin{align}
\sigma_t &= \sigma_t^\text{av} + \sigma_t^\text{G} + \sigma_t^\text{nG} \qquad \text{with} \label{entropy-decomposition} \\
\sigma_t^\text{av} &= \frac{\gamma}{T} \big\Vert \av{\bm{v}}_t \big\Vert^2 \nn
\sigma_t^\text{G} &= \frac{\gamma}{m} \text{tr}\bigg( \frac{m}{T} \bm{\Xi}_t^v + \frac{T}{m} \big( \bm{\Xi}_t^{-1} \big)^v  - 2 \bm{I} \bigg) . \n
\end{align}
The first term is obviously positive.
The third term, $\sigma_t^\text{nG}$ is the positive non-Gaussian contribution introduced in \eqref{shannon-decomposition}.
For the second term, we have
\begin{align}
\sigma_t^\text{G} &\geq \frac{\gamma}{m} \text{tr}\bigg( \frac{m}{T} \bm{\Xi}_t^v + \frac{T}{m} \big( \bm{\Xi}_t^v \big)^{-1}  - 2 \bm{I} \bigg) \\
&= \frac{\gamma}{m} \text{tr}\Bigg( \bigg( \sqrt{\frac{m}{T} \bm{\Xi}_t^v} - \sqrt{\frac{T}{m} \big( \bm{\Xi}_t^v \big)^{-1}} \bigg)^2 \Bigg) \geq 0 \n ,
\end{align}
so that this term is likewise positive.
\eqref{entropy-decomposition} gives a decomposition of the total entropy production rate into three positive contributions.
The first contribution stems from a non-zero average velocity.
The second contribution, which we refer to as the Gaussian contribution $\sigma_t^\text{G}$ only involves second moments of velocity and position.
It is therefore insensitive to whether the probability distribution of the system is Gaussian or not and quantifies the degree to which the Gaussian part of the fluctuations is out of equilibrium.
The third contribution $\sigma_t^\text{nG}$, on the other hand, vanishes when the joint distribution is Gaussian and therefore quantifies the non-Gaussian features of the system.
The presence of this term implies that a system with non-Gaussian velocity statistics necessarily exhibits a finite amount of dissipation.

Another decomposition can be obtained by using the chain rule for the Fisher information
\begin{align}
\bm{F}_t^v &= \bm{F}_t^{x \vert v} + \bmc{F}_t^v \quad \text{with} \\
\big(\bm{F}_t^{x \vert v} \big)_{kl} &= \Av{\partial_{v_k} \ln p_t^{x \vert v} \partial_{v_l} \ln p_t^{x \vert v}}_t, \nn
\big(\bmc{F}_t^{v} \big)_{kl} &= \Av{\partial_{v_k} \ln p_t^{v} \partial_{v_l} \ln p_t^{v}}_t \n .
\end{align}
Here, $\bm{F}_t^{x \vert v}$ is the Fisher information matrix of the conditional probability density of the position and $\bmc{F}_t^v$ the Fisher information matrix of the velocity probability density.
Using this, we have
\begin{align}
\sigma_t &= \sigma_t^\text{av} + \sigma_t^{\text{G},v} + \sigma_t^{\text{nG},v} + \sigma_t^{x \vert v} \quad \text{with} \label{entropy-decompositon-v} \\
\sigma_t^{\text{G},v} &= \frac{\gamma}{m} \text{tr}\bigg( \frac{m}{T} \bm{\Xi}_t^v + \frac{T}{m} \big( \bm{\Xi}_t^v \big)^{-1}  - 2 \bm{I} \bigg) \nn
\sigma_t^{\text{nG},v} &=  \frac{\gamma T}{m^2} \text{tr} \Big( \bmc{F}_t^v - \big(\bm{\Xi}_t^{v} \big)^{-1} \Big) \nn
\sigma_t^{x \vert v} &= \frac{\gamma T}{m^2} \text{tr} \big( \bm{F}_t^{x \vert v} \big) . \n
\end{align}
Applying \eqref{fisher-covariance-inequality-xv} to the probability density of the velocity, we have
\begin{align}
\bmc{F}_t^v \geq \big(\bm{\Xi}_t^v \big)^{-1} ,
\end{align}
with equality if the velocity density is Gaussian.
Since $\bm{F}_t^{x \vert v}$ is positive semidefinite, all four terms in \eqref{entropy-decompositon-v} are positive.
Similar to \eqref{entropy-decomposition}, the second and third terms represent a contribution from the Gaussian fluctuations and from a non-Gaussian probability density, respectively.
The difference is that, in \eqref{entropy-decompositon-v}, these terms only involve the statistics of the velocity instead of the joint statistics of position and velocity.
To compensate for this, the fourth term quantifies the amount of correlations between position and velocity.
The decomposition \eqref{entropy-decompositon-v} shows that the system is out of equilibrium when any one of the following conditions is not satisfied
\begin{enumerate}
\item The average velocity is zero, $\av{\bm{v}}_t = 0$.
\item The velocity fluctuations are thermal, $\bm{\Xi}_t^v = \frac{T}{m} \bm{I}$.
\item The velocity density is Gaussian.
\item The joint density factorizes, $p_t(\bm{x},\bm{v}) = p_t(\bm{x}) p_t(\bm{v})$.
\end{enumerate}
All of these conditions are necessary for thermal equilibrium and \eqref{entropy-decompositon-v} states that their violations give rise to separate, positive contributions to the entropy production rate.
For the rate of entropy change of the system, \eqref{entropy-decompositon-v} yields
\begin{align}
\sigma_t^\text{sys} &= \frac{\gamma}{m} \text{tr}\bigg(\frac{T}{m} \big(\bm{\Xi}_t^{v} \big)^{-1} - \bm{I} \bigg) + \sigma_t^{\text{nG},v} + \sigma_t^{x \vert v} \quad \text{with} \label{shannon-decomposition-v} \\
\sigma_t^{\text{nG},v} &= \frac{\gamma T}{m^2} \text{tr}\Big( \bmc{F}_t^v - \big(\bm{\Xi}_t^{v} \big)^{-1} \Big) \nn
\sigma_t^{x \vert v} &= \frac{\gamma T}{m^2} \text{tr} \big( \bm{F}_t^{x \vert v} \big) \n .
\end{align}
As a variation of \eqref{entropy-decompositon-v}, we can also write
\begin{align}
\sigma_t &= \sigma_t^\text{av} + \sigma_t^{\text{G},v} + \sigma_t^{\text{G},x \vert v} + \sigma_t^{\text{nG}} \quad \text{with} \label{entropy-decompositon-v2} \\
\sigma_t^{\text{G}, x \vert v} &= \sigma_t^{\text{G}} - \sigma_t^{\text{G},v} = \frac{\gamma T}{m^2} \text{tr}\Big(\big( \bm{\Xi}_t^{-1} \big)^{v} - \big( \bm{\Xi}_t^v \big)^{-1}\Big) \n .
\end{align}
The difference is that, in \eqref{entropy-decompositon-v}, we evaluate the non-Gaussian contribution based on the velocity statistics and the correlation contribution based on the joint statistics, whereas in \eqref{entropy-decompositon-v2}, the correlation contribution only contains the Gaussian part and the non-Gaussian contribution is evaluated based on the joint statistics.
The advantage of this decomposition is that the correlation contribution has an explicit expression in terms of the covariance matrix,
\begin{align}
\sigma_t^{\text{G}, x \vert v} = \frac{\gamma T}{m^2} \text{tr}\Big(\big( \bm{\Xi}_t^v - \bm{\Xi}_t^{xv,\text{T}} (\bm{\Xi}_t^x)^{-1} \bm{\Xi}_t^{xv} \big)^{-1} - \big( \bm{\Xi}_t^v \big)^{-1}\Big),
\end{align}
and thus three of the four positive contributions can be evaluated using only the first and second order statistics of the system.

\subsection{Bounds on entropy production rate} \label{sec-bounds}
Neglecting the positive terms in \eqref{shannon-decomposition} and \eqref{shannon-decomposition-v} yields lower bounds on the rate of entropy change,
\begin{align}
\sigma_t^\text{sys} &\geq \frac{\gamma}{m} \Bigg( \frac{T}{m} \text{tr}\bigg( \Big( \bm{\Xi}_t^v - \bm{\Xi}_t^{xv,\text{T}} \big( \bm{\Xi}_t^x \big)^{-1} \bm{\Xi}_t^{xv} \Big)^{-1} \bigg) - d \Bigg) \nn
&\geq \frac{\gamma}{m} \bigg( \frac{T}{m} \text{tr}\Big( \big(\bm{\Xi}_t^{v} \big)^{-1} \Big) - d \bigg) ,
\end{align} 
where we explicitly expressed the $\bm{v}$-component of the inverse covariance matrix in terms of its components.
These bounds only involve the second moments of position and velocity, respective only velocity, and are therefore considerably easier to evaluate than the exact expression.
Note that the first inequality becomes an equality in the case of a Gaussian joint distribution and we can therefore write
\begin{align}
d_t S_t \geq d_t S_t^\text{G},
\end{align}
where $S_t^\text{G}$ is the Gibbs-Shannon entropy of a Gaussian probability density with the same covariance matrix.
This relation is at first glance surprising, given the well-known fact that the Gibbs-Shannon entropy at given covariance is maximized for a Gaussian probability density \cite{Cover2012},
\begin{align}
S_t \leq S_t^\text{G} .
\end{align}
It implies that, for generic underdamped dynamics, the rate of change of the system's entropy and its instantaneous value satisfy opposite inequalities with the corresponding Gaussian result.
Intuitively, this finding can be understood as follows:
In thermal equilibrium, the velocity statistics of the system are Gaussian and therefore have maximal entropy for a given variance of the velocity.
Any non-Gaussian statistics decrease the entropy and therefore correspond to a system that is further from thermal equilibrium compared to Gaussian statistics.
The relation $d_t S_t \geq d_t S_t^\text{G}$ states that the decreased instantaneous entropy is compensated by a faster rate of increase as the system tends towards its maximum entropy equilibrium state.
\eqref{entropy-decomposition} also implies a positive lower bound on the entropy production rate,
\begin{align}
\sigma_t &\geq \sigma_t^\text{av} + \sigma_t^{\text{G}} . \label{entropy-bound-total}
\end{align}
In terms of the components of the covariance matrix, the second term explicitly reads
\begin{align}
\sigma_t^\text{G} = \frac{\gamma}{m} \text{tr} \bigg( &\frac{T}{m} \bm{\Xi}_v^{-1}  \\
&+ \frac{m}{T} \Big( \bm{\Xi}_t^v - \bm{\Xi}_t^{xv,\text{T}} \big( \bm{\Xi}_t^x \big)^{-1} \bm{\Xi}_t^{xv} \Big)^{-1} - 2 \bm{I} \bigg) \n .
\end{align}
\eqref{entropy-bound-total} can provide a useful lower estimate on the total entropy production rate, since it only depends on the first and second moments of the probability density and is therefore straightforward to evaluate from trajectory data.
The bound is tight whenever the joint probability density is approximately Gaussian.

While the non-Gaussian contributions defined in \eqref{entropy-decomposition} and \eqref{entropy-decompositon-v} imply a positive amount of dissipation whenever the joint, respectively velocity, density are non-Gaussian, they depend on the parameters of the system, in particular the temperature, which is not straightforward to obtain from the non-equilibrium statistics of the system alone.
To obtain a bound that only depends on the statistics of the system, we start from the expression
\begin{align}
\sigma_t \geq \frac{\gamma}{m} \text{tr} \bigg( \frac{m}{T} \bm{\Xi}_t^v + \frac{T}{m} \bmc{F}_t^v - 2 \bm{I} \bigg).
\end{align}
We rewrite this as
\begin{align}
\sigma_t \geq \frac{\gamma}{m} \text{tr} \Bigg( \bigg( \sqrt{\frac{m}{T} \bm{\Xi}_t^v} - \sqrt{\frac{T}{m} \bmc{F}_t^v} \bigg)^2 + 2 \sqrt{\bm{\Xi}_t^v} \sqrt{\bmc{F}_t^v} - 2 \bm{I} \Bigg).
\end{align}
Since the first term is positive, we have
\begin{align}
\sigma_t \geq 2 \frac{\gamma}{m} \text{tr} \Big( \sqrt{\bm{\Xi}_t^v} \sqrt{\bmc{F}_t^v} - \bm{I} \Big).
\end{align}
Since the inequality between $\bmc{F}_t^v$ and $\bm{\Xi}_t^v$ translates into the same inequality between their square roots,
\begin{align}
\bmc{F}_t^v \geq \big(\bm{\Xi}_t^v \big)^{-1} \quad \Rightarrow \quad \sqrt{\bmc{F}_t^v} \geq \sqrt{\big(\bm{\Xi}_t^v \big)^{-1}},
\end{align}
the expression on the right-hand side is also positive.
We define the thermalization time $\tau_\text{th} = m/\gamma$, that is, the characteristic relaxation timescale of the velocity.
Using this, we have
\begin{align}
\tau_\text{th} \sigma_t \geq 2 \text{tr} \Big( \sqrt{\bm{\Xi}_t^v} \sqrt{\bmc{F}_t^v} - \bm{I} \Big) .
\end{align}
The right-hand side vanishes in the Gaussian case.
It only depends on the velocity distribution and is independent of the parameters of the system.
This implies that we can give a lower bound on the entropy production during the characteristic thermalization time by only using the non-Gaussian features of the velocity distribution.

Finally, let us focus on the special case of a non-equilibrium steady state.
In this case, the entropy of the system remains constant and only the environmental contribution in \eqref{entropy-system-environment} remains.
Further, as we noted in \eqref{fisher-steady-temperature}, we have the identity
\begin{align}
\frac{T}{m} = \frac{d}{\text{tr}\big( \bm{F}_\text{st}^v \big)} .
\end{align}
This allows us to write the environmental entropy production as
\begin{align}
\sigma_\text{st}^\text{env} = \sigma_\text{st}^\text{av} + \frac{\gamma}{d m} \Big( \text{tr}\big( \bm{\Xi}_\text{st}^v \big) \text{tr} \big( \bm{F}_\text{st}^v \big) - d^2 \Big) .
\end{align}
Note that the right-hand side is explicitly positive since $\text{tr}(\bm{F}_\text{st}^v) \geq \text{tr}( (\bm{\Xi}_\text{st}^v)^{-1}) \geq d^2/\text{tr}(\bm{\Xi}_\text{st}^v)$.
In particular, using the first inequality, we have the non-zero lower bound on the steady state entropy production rate
\begin{align}
\sigma_\text{st}^\text{env} = \sigma_\text{st}^\text{av} +\frac{\gamma}{d m} \Big( \text{tr}\big( \bm{\Xi}_\text{st}^v \big) \text{tr} \big( (\bm{\Xi}_\text{st}^v)^{-1} \big) - d^2 \Big),
\end{align}
which only involves the first and second order moments of the velocity.
In one dimension, the second term vanishes, but in higher dimensions, it is non-zero whenever the covariance matrix of the velocity is not proportional to the identity matrix, i.~e.~the distribution of the system does not correspond to a thermal distribution at some effective temperature.
By using \eqref{fisher-variational-2}, we can obtain tighter bounds on the Fisher information by specific choices for the function $\phi(\bm{x},\bm{v})$.
Writing
\begin{align}
\phi(\bm{x},\bm{v}) = \frac{c_1}{2} &\big( \bm{v} - \Av{\bm{v}}_\text{st} \big)^\text{T} \big( \bm{\Xi}_\text{st}^v \big)^{-1} \big( \bm{v} - \Av{\bm{v}}_\text{st} \big) \\
& + c_2 \psi(\bm{x},\bm{v}) \n ,
\end{align}
and optimizing with respect to $c_1$ and $c_2$, we obtain
\begin{align}
\text{tr} &\big( \bm{F}_\text{st}^v \big) \geq \text{tr}\big( (\bm{\Xi}_\text{st}^v)^{-1} \big) \Bigg( 1 + \\
& \frac{\big(\Av{\grad_v^2 \psi}_\text{st} - \Av{\grad_v \psi (\bm{\Xi}_\text{st}^v)^{-1} (\bm{v} - \av{\bm{v}}_\text{st} }_\text{st}\big)^2}{\text{tr}\big( (\bm{\Xi}_\text{st}^v)^{-1} \big) \Av{\Vert \grad_v \psi \Vert^2}_\text{st} - \Av{\grad_v \psi (\bm{\Xi}_\text{st}^v)^{-1} (\bm{v} - \av{\bm{v}}_\text{st} }_\text{st}^2 } \Bigg) \n .
\end{align}
This gives a tighter lower bound on the Fisher information (compared to the inverse covariance matrix) for any function $\psi(\bm{x},\bm{v})$.

\subsection{Explicit bounds for one-dimensional systems} \label{sec-bounds-1D}
While in higher dimensions, the number of necessary basis functions is generally large and determining their optimal coefficients according to \eqref{entropy-system-estimate} is only feasible numerically, more explicit results can be obtained in one dimension.
We set $\psi_q(v) = v^q$ where $q \in M$ with $M$ a set of integers.
Depending on the choice of $M$, we can obtain different bounds on the Fisher information.
Some non-trivial examples are
\begin{align}
\mu_2 F_t^v \geq \left\lbrace \begin{array}{ll}
1 &\text{for} \;  M = \lbrace 1, 2 \rbrace \\[1ex]
\frac{9 \mu_2^3}{\mu_6 - \mu_3^2} &\text{for} \;  M = \lbrace 1, 4 \rbrace \\[1ex]
1 + \frac{\mu_3^2}{\mu_4 \mu_2 - \mu_3^2} &\text{for} \;  M = \lbrace 2, 3 \rbrace \\[1ex]
1 + \frac{(\mu_4- 3 \mu_2)^2}{\mu_6 \mu_2 - \mu_4^2} &\text{for} \;  M = \lbrace 2, 4 \rbrace \\[1ex]
1 + \frac{\mu_3^2}{\mu_4 \mu_2 - \mu_3^2 - \mu_2^2} &\text{for} \;  M = \lbrace 1, 2, 3 \rbrace \\[1ex]
1 + \frac{(\mu_4- 3 \mu_2)^2}{\mu_6 \mu_2 - \mu_3^2 \mu_2 - \mu_4^2} &\text{for} \;  M = \lbrace 1, 2, 4 \rbrace ,
\end{array} \right.
\end{align}
where $\mu_n$ is the $n$-th central moment of the velocity, $\mu_n = \Av{(v-\Av{v}_t)^n}_t$.
The first bound represents the well-known relation between the Fisher information and the variance, see \eqref{fisher-covariance}.
The following lines represent tighter lower bounds on the Fisher information in terms of higher-order moments.
Note that the bounds get tighter as we include more basis functions, for example the bound for $M = \lbrace 1,2,4 \rbrace$ is tighter than either one for $M = \lbrace 1,2 \rbrace$ and for $M = \lbrace 1,4 \rbrace$, while the relation between the latter two depends on the concrete probability density.
We also see that including basis functions up to $v^N$ results in bounds involving moments up to order $\mu_{2N-2}$.
Any lower bound on the Fisher information also results in a lower bound on the rate of entropy change. 
We have, for example,
\begin{align}
\sigma_t^\text{sys} \geq \frac{\gamma}{m} \Bigg( \frac{T}{m \mu_2} \bigg( 1 + \frac{(\mu_4- 3 \mu_2)^2}{\mu_6 \mu_2 - \mu_3^2 \mu_2 - \mu_4^2} \bigg) - 1 \Bigg) ,
\end{align}
which provides a concrete lower bound on the rate of entropy change in terms of the second, third, fourth and sixth moment of the velocity density.

\section{Description of the experiment} \label{sec-experiment}
Here we briefly discuss the experimental apparatus used to provide experimental data that could be analyzed as described in the previous section. \\ 

\begin{figure*}
\includegraphics[width=15cm]{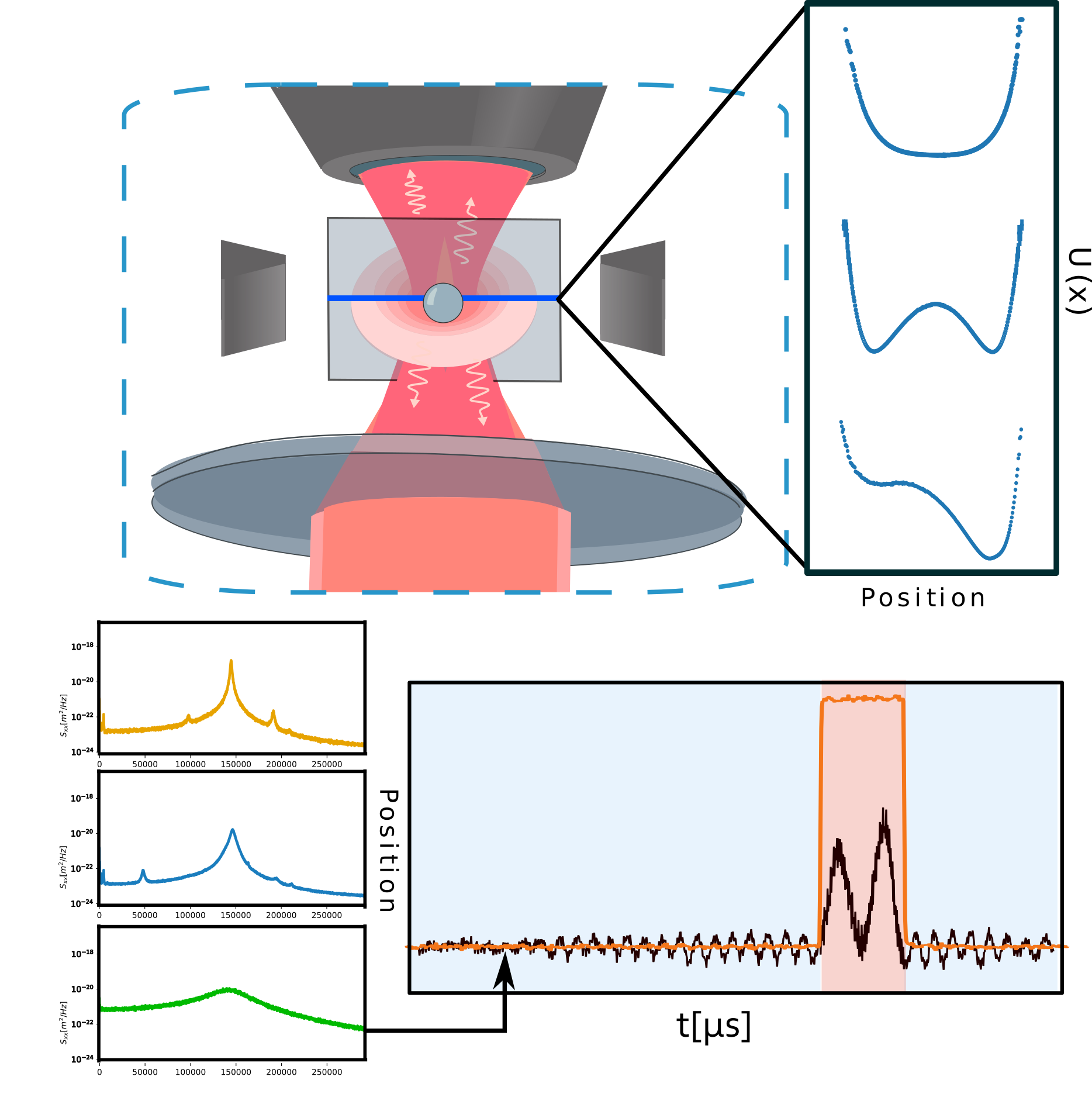}
\caption{(a) Schematic representation of the experimental setup to generate non-linear potentials. (b) Representations of the 3 kinds of potential generated in the pulsed protocols. (c) Motional spectra for the different initial states, at 2mBar without and with pre-cooling, and at 80mBar. (d) Example protocol with 9$\mu s$ evolution in a quartic potential.}
\label{Setup}
\end{figure*}

The system depicted schematically in fig.\ref{Setup}(a), allows the generation of an optical trap with tunable optical potentials in the focus of a microscope objective (CFI TU Plan Fluor EPI 50x, Nikon Corp., numerical aperture NA = 0.8). This is achieved by overlapping a trapping beam with a spatially shaped beam controlled using a spatial light modulator (SLM, Meadowlark Optics 512x512 pixels). Using acousto-optical modulators, the intensities of both beams are controllable on MHz timescales, allowing fast switching between different optical potentials.\\ 
The shaped beam enables the generation of complex potential landscapes along one spatial axis, here defined as the x-axis. By utilizing charged test particles the potentials can be further modified using DC electric fields applied via a pair of electrodes aligned along the x-axis. 
The data were acquired for three different potential configurations; see Fig.\ref{Setup}(b), these being cubic ($x^3$), quartic ($x^4$) and inverted ($-x^2$) contributions.\\
In addition to this, the pressure of the vacuum system can be varied, effectively controlling the damping the system experiences, allowing both under- and overdamped dynamics to be studied.

\subsection{Experimental protocols}
In the experiment pulsed protocols were performed to measure the dynamics in the prepared potentials. This allows great control of the initial state, which can be cooled to a reduced effective temperature using cold damping \textcolor{red}{\cite{Magrini2021}}.\\
One such protocol seen in Fig.\ref{Setup}(c) consists of an approximately 20 ms period where the state is prepared in the harmonic trapping potential, followed by an evolution time $\tau^\text{pulse}=9\mu s$ in the chosen potential. This is performed approximately 10,000 times for each experimental setting, resulting in a collection of timetraces $x_i(t_j)$ with $i$ identifying the timetrace index, and $j$ the time interval. The data has been acquired with a sample rate of 4MS/s. \\
For each of the three chosen potential shapes, three experimental runs were performed. Two were performed in the under damped regime, at a pressure $P_1$ of approx. 2mBar, with two effective initial temperatures $T_{1,2}$, 300K and 66K.
The third run was performed at a pressure $P_2$ of 80mBar, close to the overdamped regime. The experimental parameters are listed in table \ref{params}.
\begin{table}
\begin{tabular}{|c| c| c| c|c|} 
 \hline
 $P_1$[mBar] & $P_2$[mBar] & $T_1$[K] & $T_2$[K] &$t_e$[$\mu s$]\\ [0.5ex] 
 \hline\hline
 2.0$\pm 0.6$ & 80$\pm 20$ & 296$\pm 2$ & 66$\pm 5$ &9.00$\pm 0.05$\\ 
 \hline

 \hline
\end{tabular}
\caption{Summary of experimental parameters.}
\label{params}
\end{table}

\section{Estimating entropy production and non-Gaussianity from experimental trajectories} \label{sec-data-analysis}
Given the experimental timetraces and the result of  Section \ref{sec-estimation-bounds}, we want to estimate the entropy production rate for levitated particles driven by transient non-linear potentials. As the system is sufficiently underdamped, the noise in the data does not prevent extracting the instantaneous velocity from differences in the position; concretely, we use the central difference $v(t) = (x(t+\Delta t) - x(t-\Delta t))/(2 \Delta t)$, where $\Delta t$ is the measurement time-resolution, see Appendix \ref{app-v-estim}. From this, we need to derive the thermal velocity $v_{th}$, the Fisher information matrix $\bm{F}_t^{v}$ and the covariance matrix $\bm{\Xi}_t^{-1}$.

\subsection{Extracting the thermal velocity and dimensionless entropy production rate}
Extracting the above parameters would in principle require the knowledge of the parameters of the system, such as damping coefficient, particle mass and temperature. However, experimentally those parameters are known only to a certain precision. The temperature experienced by the particle, in particular, may not correspond to the environmental temperature because of fluctuations in the laser power and thus the optical force exerted on the particle.
However, without knowledge of these parameters, we cannot evaluate the explicit expressions such as \eqref{entropy-system-environment}.
Specifically, the estimates and bounds for the entropy production rate derived in Section \ref{sec-estimation-bounds} depend on two parameters, the thermalization timescale $\tau_\text{th} = m/\gamma$ and the ratio of temperature and mass, or thermal velocity $v_\text{th}^2 = T/m$.
If the particle is in equilibrium at an effective temperature $T$, we can determine the ratio $T/m$ from equipartition of energy,
\begin{align}
\Av{v_i^2}_\text{eq} = v_\text{th}^2 = \frac{T}{m} \label{equipartition} .
\end{align}
However, according to \eqref{fisher-steady-temperature}, a more general expression, also valid in non-equilibrium steady states, is given by
\begin{align}
v_\text{th}^2 = \frac{d}{\text{tr}\big(\bm{F}_\text{st}^v\big)} \label{thermal-fisher} .
\end{align}
This allows us to determine the thermal velocity irrespective of whether the system is in thermal equilibrium or not.
Moreover, if the two expressions differ, this provides an indication that the system is in a non-equilibrium steady state.
For a freely diffusing particle, the thermalization time $\tau_\text{th}$ can be obtained from the decay of the velocity-autocorrelation function.
However, in the presence of other forces, the decay rate of the correlation function is generally modified.
For a harmonic trapping force $f(x) = -\kappa x$ and in thermal equilibrium, another way of obtaining the thermalization time is to compute the power-spectral density of the velocity,
\begin{align}
S_v(\omega) = \frac{\gamma T \omega^2}{(k - m \omega^2)^2 + \gamma^2 \omega^2} .
\end{align}
It is straightforward to see that this quantity is maximal at the natural frequency of the oscillator $\omega_0 = \sqrt{\kappa/m}$, where its value is given by
\begin{align}
S_v\bigg(\sqrt{\frac{\kappa}{m}} \bigg) = \frac{T}{\gamma} = v_\text{th}^2 \tau_\text{th} .
\end{align}
Thus, knowing the thermal velocity, we can read off the thermalization time from the peak value of the power-spectral density.
However, just like \eqref{equipartition}, this result relies crucially on thermal equilibrium and in addition on the linearity of the trapping force.
For a system in a non-equilibrium state and/or subject to non-linear forces, it is therefore not straightforward to estimate the thermalization time $\tau_\text{th}$ from the data.
One way around this is to note that the various relations for the entropy production rate derived in Section \ref{sec-estimation-bounds} only depend on the thermalization time via an overall prefactor.
Thus, if we define the dimensionless entropy production rate
\begin{align}
\bar{\sigma}_t = \tau_\text{th} \sigma_t,
\end{align}
then the corresponding results become independent of the thermalization time and only depend on the thermal velocity, which we can obtain from the data using \eqref{thermal-fisher}.
Intuitively, $\bar{\sigma}_t$ measures the amount entropy production during the characteristic thermalization time.
Focusing in particular on the decomposition \eqref{entropy-decompositon-v2}, we have
\begin{align}
\bar{\sigma}_t &= \bar{\sigma}_t^\text{av} + \bar{\sigma}_t^{\text{G},v} + \bar{\sigma}_t^{\text{G},x \vert v} + \bar{\sigma}_t^{\text{nG}} \quad \text{with} \label{entropy-decompositon-dimensionless} \\
&\bar{\sigma}_t^\text{av} = \frac{1}{v_\text{th}^2} \Vert \av{\bm{v} }\Vert^2 \nn
&\bar{\sigma}_t^{\text{G},v} = \text{tr} \bigg( \frac{1}{v_\text{th}^2} \bm{\Xi}_t^{v} + v_\text{th}^2 \big( \bm{\Xi}_t^{v} \big)^{-1} \bigg) - 2 d \nn
&\bar{\sigma}_t^{\text{G}, x \vert v} = v_\text{th}^2 \text{tr}\Big(\big( \bm{\Xi}_t^{-1} \big)^{v} - \big( \bm{\Xi}_t^{v} \big)^{-1}\Big) \nn
&\bar{\sigma}_t^{\text{nG}} =  v_\text{th}^2 \text{tr}\Big( \bm{F}_t^{v} - \big(\bm{\Xi}_t^{-1} \big)^{v} \Big) \n .
\end{align}
We stress that all quantities apart from $v_\text{th}$ can be determined solely from a measurement of the instantaneous velocity statistics; determining $v_\text{th}$ additionally requires performing a measurement in a (not necessarily equilibrium) steady state.
For the dimensionless rate of Shannon entropy change we obtain
\begin{align}
\bar{\sigma}^\text{sys}_t &= \bar{\sigma}_t^{\text{sys},\text{G}} + \bar{\sigma}_t^{\text{nG}} \quad \text{with} \label{shannon-decompositon-dimensionless} \\
&\bar{\sigma}_t^\text{sys,G} = v_\text{th}^2 \text{tr} \big(  ( \bm{\Xi}_t^{v} )^{-1} \big) -  d \n .
\end{align}
Note that, since the first term in $\bar{\sigma}_t^\text{sys,G}$ is positive, we have universal lower bound $\bar{\sigma}_t^\text{sys} \geq \bar{\sigma}_t^\text{sys,G} \geq -d$.
Since a decrease in the Shannon entropy corresponds to a contraction of the probability density, this implies that the maximal rate of contraction is bounded by the inverse thermalization time.
This maximally negative rate of Shannon entropy change is realized by an infinitely broad but Gaussian velocity density.

\subsection{Estimation of Fisher information}
While the covariance matrix entering the Gaussian contributions in \eqref{entropy-decompositon-dimensionless} can be straightforwardly determined using well-established covariance estimators, for the non-Gaussian contribution we estimate the trace of the Fisher information matrix using the variational expression \eqref{fisher-variational}.
We focus on the one-dimensional case and parameterize the variational function as
\begin{align}
\phi(x,v) = \sum_{n_x + n_v \leq M} c_{n_x,n_v} H_{n_x}(x) H_{n_v}(v),
\end{align}
where $n_x,n_v \geq 0$ are natural numbers and $H_n(x)$ denotes the Hermite polynomial of order $n$, defined as
\begin{align}
H_n(x) = (-1)^n e^{\frac{x^2}{2}} d_x^n e^{-\frac{x^2}{2}} ,
\end{align}
so that the function $\phi(x,v)$ is a polynomial of order $M$ in $x$ and $v$.
As we saw in Section~\ref{sec-bounds-1D}, the corresponding estimate on the Fisher information contains moments $x$ and $v$ up to order $2M-2$.
In principle, the estimate \eqref{fisher-estimate} of the Fisher information becomes more accurate as $M$ increases.
However, this is only true as long as we can reliably estimate the corresponding moments from the data; in particular, higher-order moments generally require better statistics.

Another issue is that, while the expression on the right-hand side of \eqref{fisher-estimate} is a lower bound on the Fisher information, this is only true if we have an infinite number of samples allowing us to exactly compute $\bm{b}$ and $\bm{D}$.
Concretely, suppose that we have a set of $N$ independent measurements $(x_k,v_k)$, $k = 1,\ldots,N$.
From these measurements and for a given set of basis functions, we can estimate the vector $\bm{b}$ on right-hand side of \eqref{fisher-estimate} using
\begin{align}
\big(\bm{b} \big)_q \estimates \big(\hat{\bm{b}}\big)_q = \frac{1}{N} \sum_{k = 1}^N \grad_v^2 \psi_q(x_k,v_k) , 
\end{align}
and similar for the estimate of the matrix $\bm{D} \estimates \hat{\bm{D}}$ from the data.
We might then estimate the right-hand side of \eqref{fisher-estimate} by
\begin{align}
\bm{b} \cdot \bm{D}^{-1} \bm{b} \estimates \hat{\bm{b}} \cdot \hat{\bm{D}}^{-1} \hat{\bm{b}} \label{fisher-bound-estimate} .
\end{align}
However, since this is a convex function of the entries of $\hat{\bm{b}}$ and $\hat{\bm{D}}$, we also have from Jensen's inequality,
\begin{align}
\Av{\hat{\bm{b}} \cdot \hat{\bm{D}}^{-1} \hat{\bm{b}}} \geq \av{\hat{\bm{b}}} \cdot \av{\hat{\bm{D}}}^{-1} \av{\hat{\bm{b}}} = \bm{b} \cdot \bm{D}^{-1} \bm{b} .
\end{align}
Thus, on average, the estimate obtained from a finite data set over-estimates the true value; in other words, the estimate is biased.
This also means that, while the true value $\bm{b} \cdot \bm{D}^{-1} \bm{b}$ is a lower bound on the Fisher information, the same is generally not true for the estimate obtained from a finite set of measurement data.
An extreme example illustrating this is when the number of basis function exceeds the number of data points, in which case the matrix $\bm{D}$ becomes singular and thus the estimate of the right-hand side of \eqref{fisher-estimate} diverges.
This also highlights the trade-off between the available data and the number of basis functions in order to accurately estimate the Fisher information.

\begin{figure}
\includegraphics[width=\linewidth]{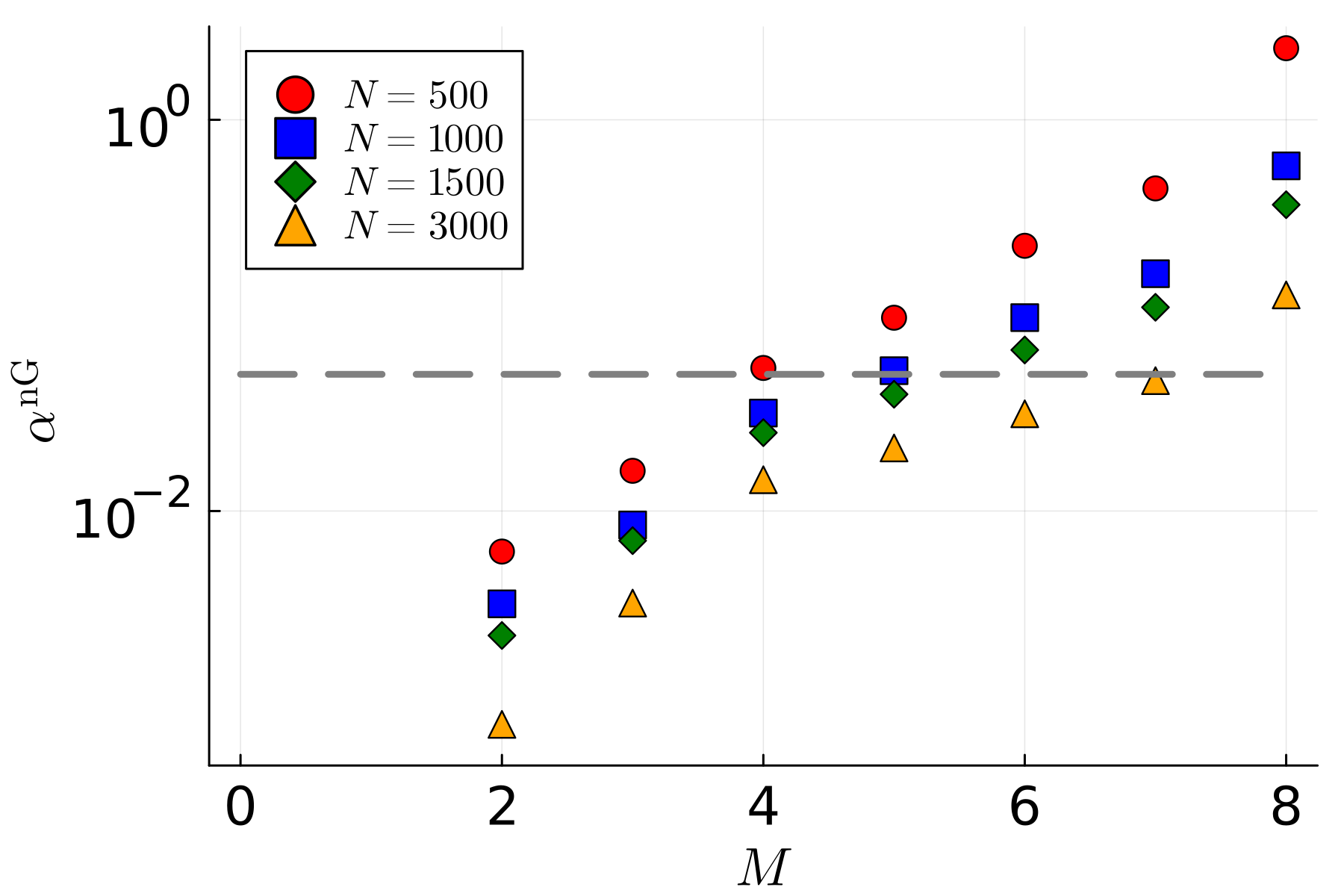}
\caption{The non-Gaussian parameter \eqref{non-gaussian-1d} as a function of the polynomial order $M$ of the basis functions for different sizes $N$ of the data set. The dashed line corresponds to a value of $\alpha^\text{nG} = 0.05$.}
\label{fig-fisher-calibration}
\end{figure}

In practice, we use $M = 4$ which corresponds to 15 basis functions and an estimate for the Fisher information involving moments of $x$ and $v$ of up to 6th order.
To justify this, consider
\begin{align}
\alpha^\text{nG} = \frac{1}{d} \text{tr}\big( \bm{F}_t^v \bm{\Xi}_t^v - \bm{I} \big) \label{non-gaussian},
\end{align}
Since $\bm{F}_t^v \geq (\bm{\Xi}_t^v)^{-1}$ with equality in equilibrium, where the velocity density becomes Gaussian, $\alpha^\text{nG}$ is a positive, dimensionless measure of the non-Gaussian nature of the velocity distribution.
In the one-dimensional case, it can be written as
\begin{align}
\alpha^\text{nG} = \frac{F_t^v - \frac{1}{\Av{v^2}}}{\frac{1}{\Av{v^2}}} \label{non-gaussian-1d}.
\end{align}

In particular, if the system is in equilibrium, then we expect $\alpha^\text{nG} = 0$.
In order to determine the accuracy of the estimate of the Fisher information, we estimate the right-hand side of \eqref{fisher-estimate} for calibration data of a trapped particle in thermal equilibrium and use the result as an estimate for the Fisher information in \eqref{non-gaussian-1d}.
The results are shown in Fig.~\ref{fig-fisher-calibration}.
Since in equilibrium, we should have $\alpha^\text{nG} = 0$, a finite value of $\alpha^\text{nG}$ stems from the error of estimating it from a finite data set.
This is confirmed by Fig.~\ref{fig-fisher-calibration}, where the estimate of $\alpha^\text{nG}$ is seen to decrease monotonically when increasing the size of the data set $N$.
On the other hand, the estimate also increases monotonically when increasing the polynomial order $M$ of the basis functions, as higher-order moments require more data for precise estimates.
Since \eqref{non-gaussian-1d} is the relative deviation of the Fisher information from its Gaussian value, it therefore directly quantifies the relative error of the estimate in the equilibrium case.
In order to have a relative deviation of less than $0.05$ for a sample size of $N = 1000$, which is the typical sample size for the following discussion, we therefore can use basis functions up to order $M = 4$.

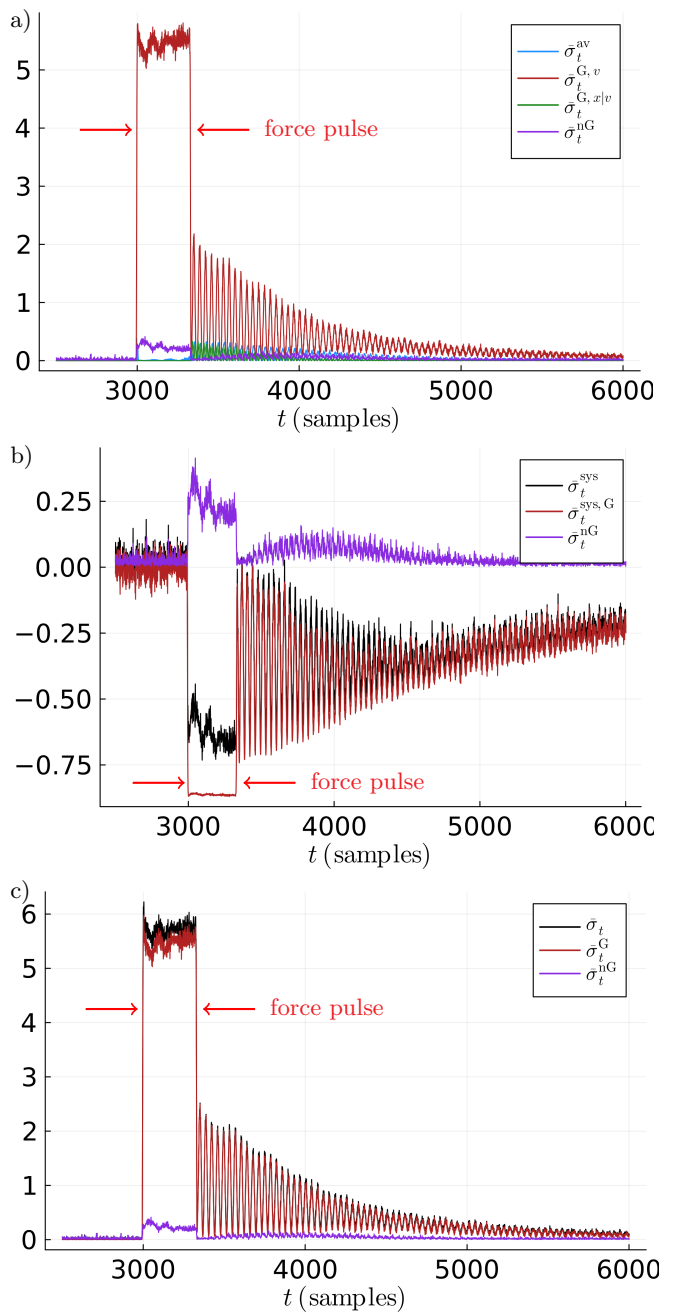
\begin{figure}[h!]
    \centering
    \begin{tikzpicture}
        \node[anchor=south west,inner sep=0] (image) at (0,0) {\includegraphics[width=\linewidth]{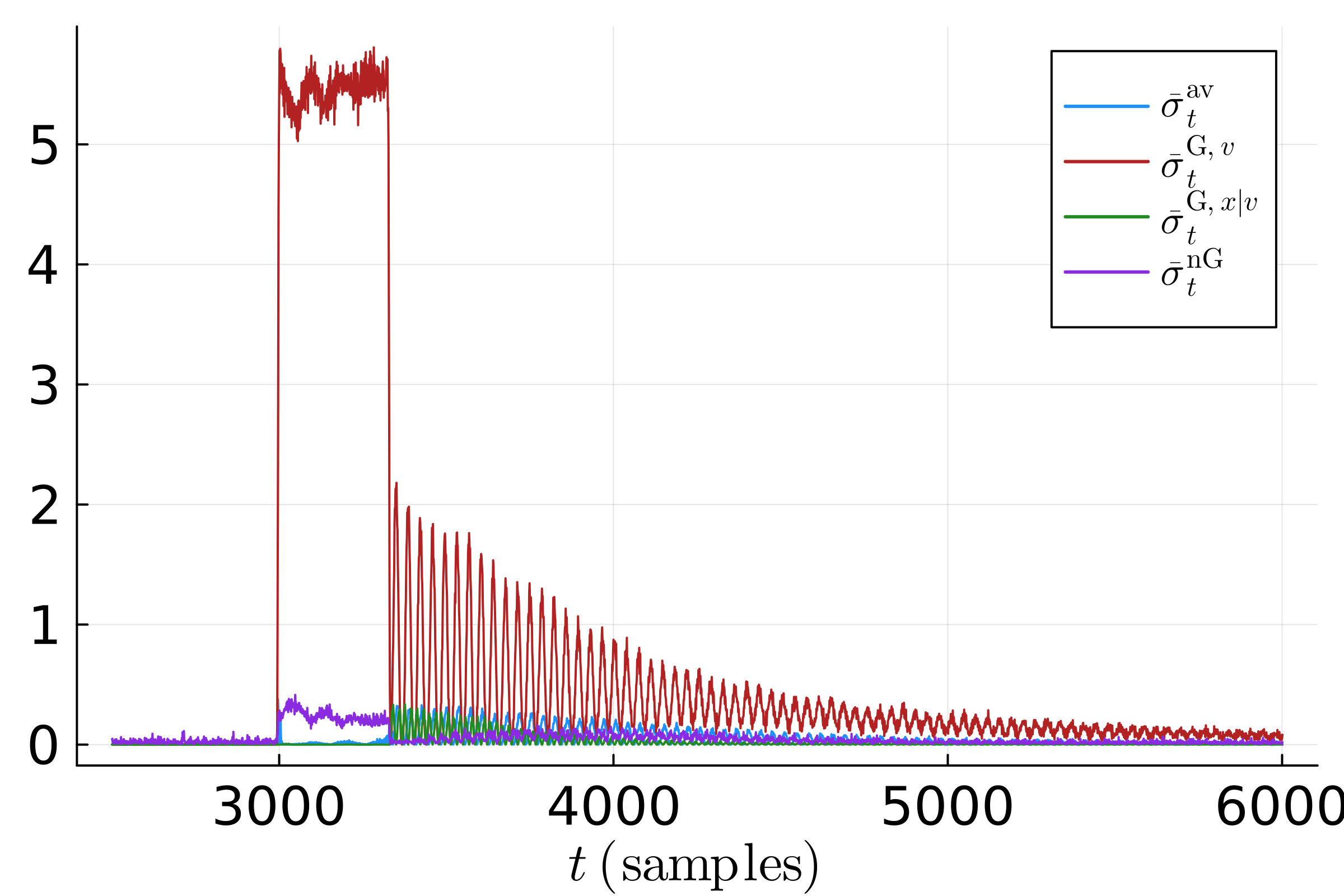}};
        \begin{scope}[x={(image.south east)},y={(image.north west)}]
            \node at (0.03,0.95) {a)};
            \draw[red,thick, ->] (0.12,0.7) -- (0.2,0.7);
            \draw[red,thick, <-] (0.3,0.7) -- (0.38,0.7);
            \node[anchor=west] at (0.39,0.7) {\Red{force pulse}};
        \end{scope}
    \end{tikzpicture}
    \begin{tikzpicture}
        \node[anchor=south west,inner sep=0] (image) at (0,0) { \includegraphics[width=\linewidth]{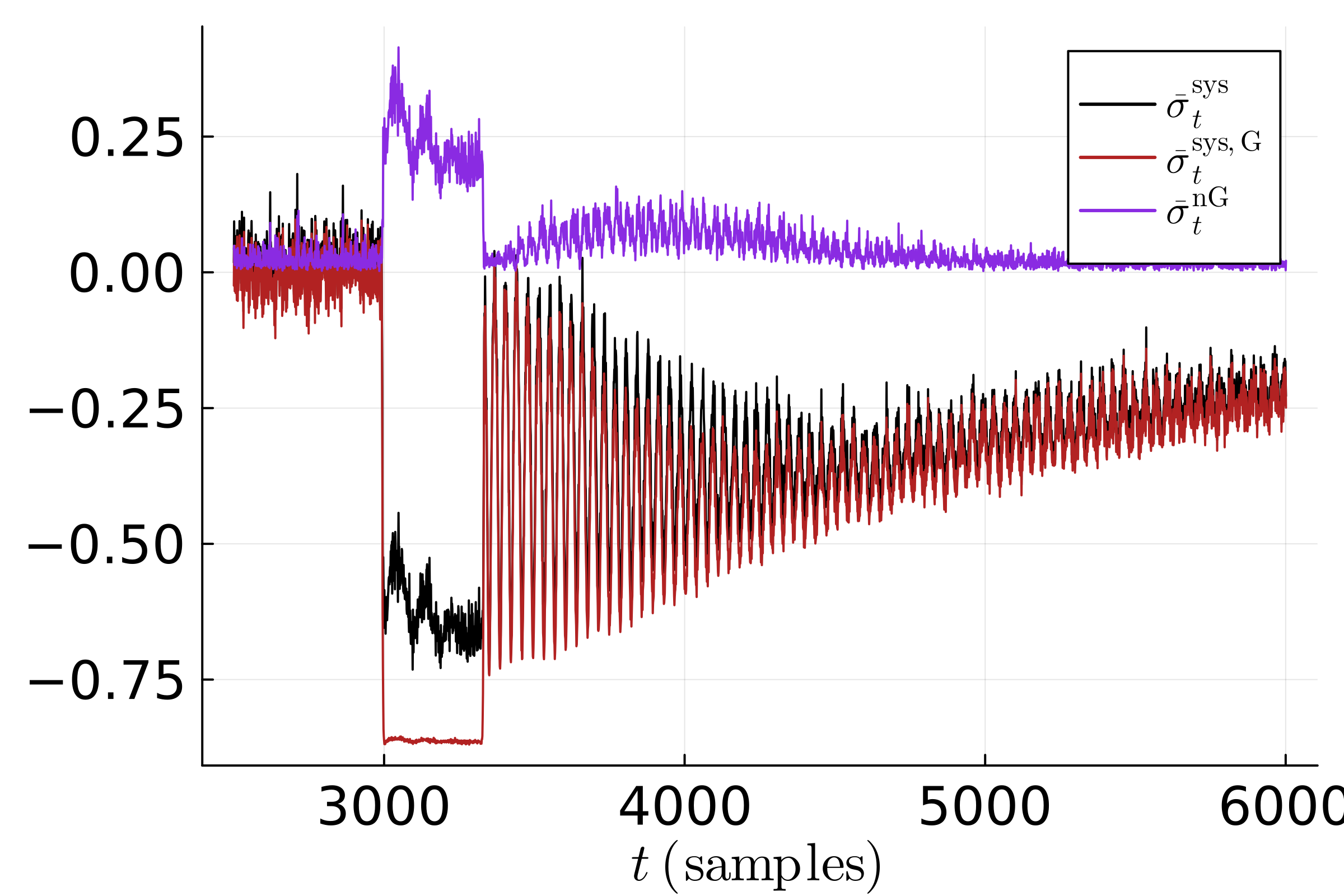}};
        \begin{scope}[x={(image.south east)},y={(image.north west)}]
            \node at (0.03,0.95) {b)};
            \draw[red,thick, ->] (0.2,0.2) -- (0.28,0.2);
            \draw[red,thick, <-] (0.37,0.2) -- (0.45,0.2);
            \node[anchor=west] at (0.46,0.2) {\Red{force pulse}};
        \end{scope}
    \end{tikzpicture}
    \begin{tikzpicture}
        \node[anchor=south west,inner sep=0] (image) at (0,0) {\includegraphics[width=\linewidth]{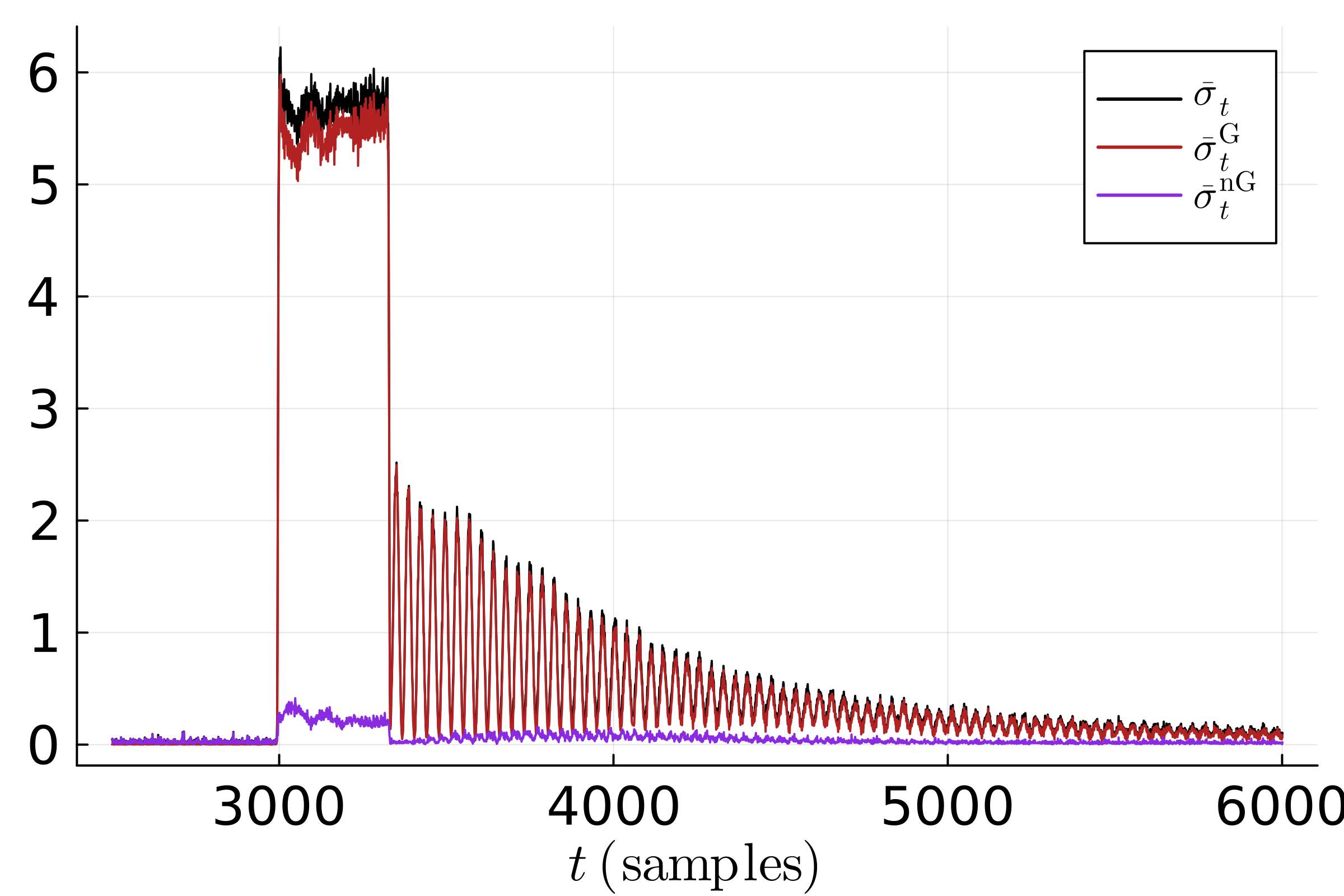}};
        \begin{scope}[x={(image.south east)},y={(image.north west)}]
            \node at (0.02,0.97) {c)};
            \draw[red,thick, ->] (0.12,0.7) -- (0.2,0.7);
            \draw[red,thick, <-] (0.3,0.7) -- (0.38,0.7);
            \node[anchor=west] at (0.39,0.7) {\Red{force pulse}};
        \end{scope}
    \end{tikzpicture}
    \caption{a) The four positive contributions to the dimensionless entropy production rate \eqref{entropy-decompositon-dimensionless} as a function of time for a trapped particle driven by a quartic-potential force pulse.
    b) The two contributions to the dimensionless rate of change of the Shannon entropy \eqref{shannon-decompositon-dimensionless} and the total value as a function of time.
    c) The overall Gaussian and non-Gaussian contribution to the entropy production rate as a function of time.
    The force pulse starts at $t = 3000$ samples and lasts $375$ samples. The sampling rate is $4.17 \cdot 10^6$ samples/s.}
    \label{fig-entropy-decop-quartic}
\end{figure}

\begin{figure}[h!]
    \centering
    \begin{tikzpicture}
        \node[anchor=south west,inner sep=0] (image) at (0,0) {\includegraphics[width=\linewidth]{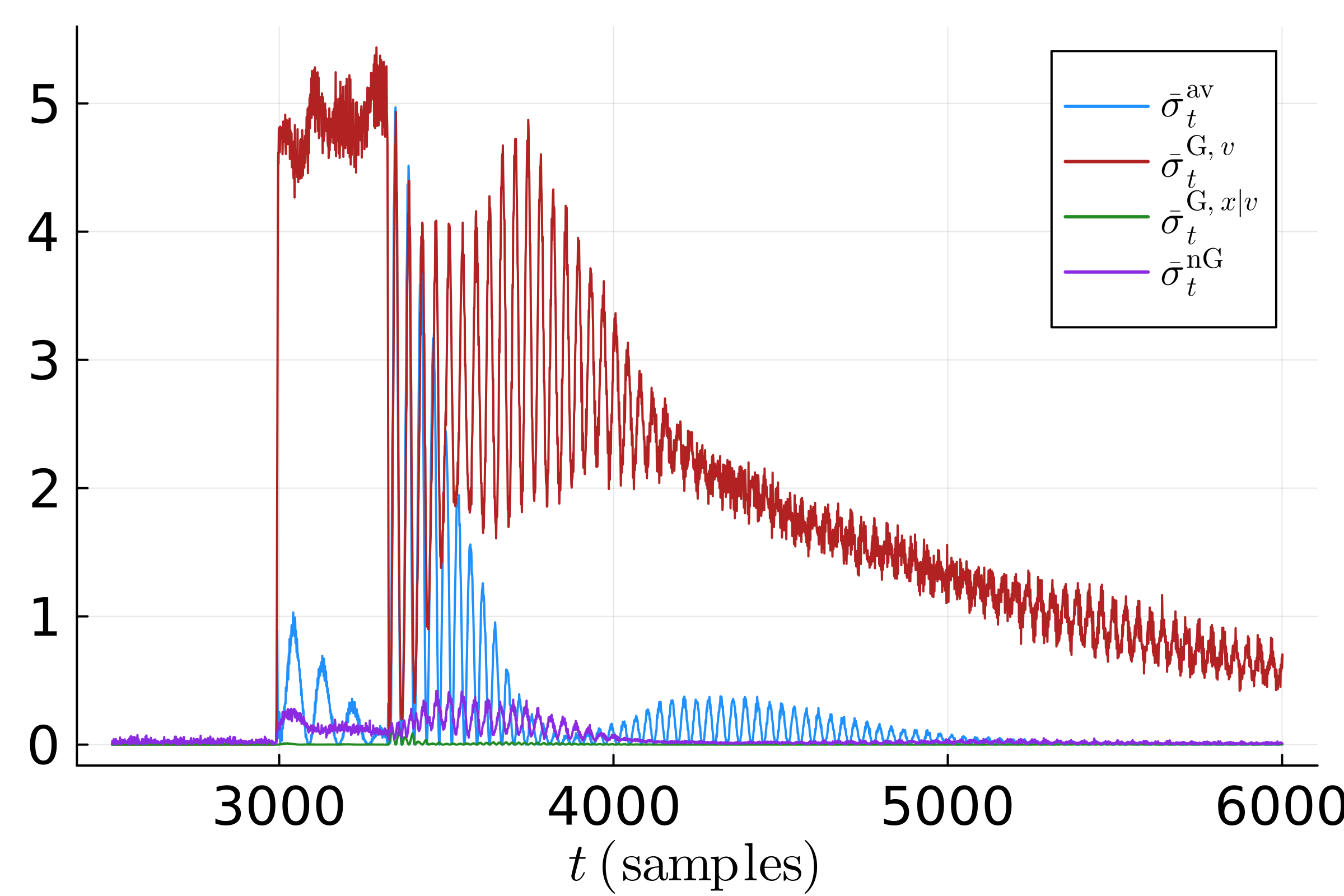}};
        \begin{scope}[x={(image.south east)},y={(image.north west)}]
            \node at (0.03,0.95) {a)};
            \draw[red,thick, ->] (0.12,0.9) -- (0.2,0.9);
            \draw[red,thick, <-] (0.29,0.9) -- (0.37,0.9);
            \node[anchor=west] at (0.38,0.9) {\Red{force pulse}};
        \end{scope}
    \end{tikzpicture}
    \begin{tikzpicture}
        \node[anchor=south west,inner sep=0] (image) at (0,0) { \includegraphics[width=\linewidth]{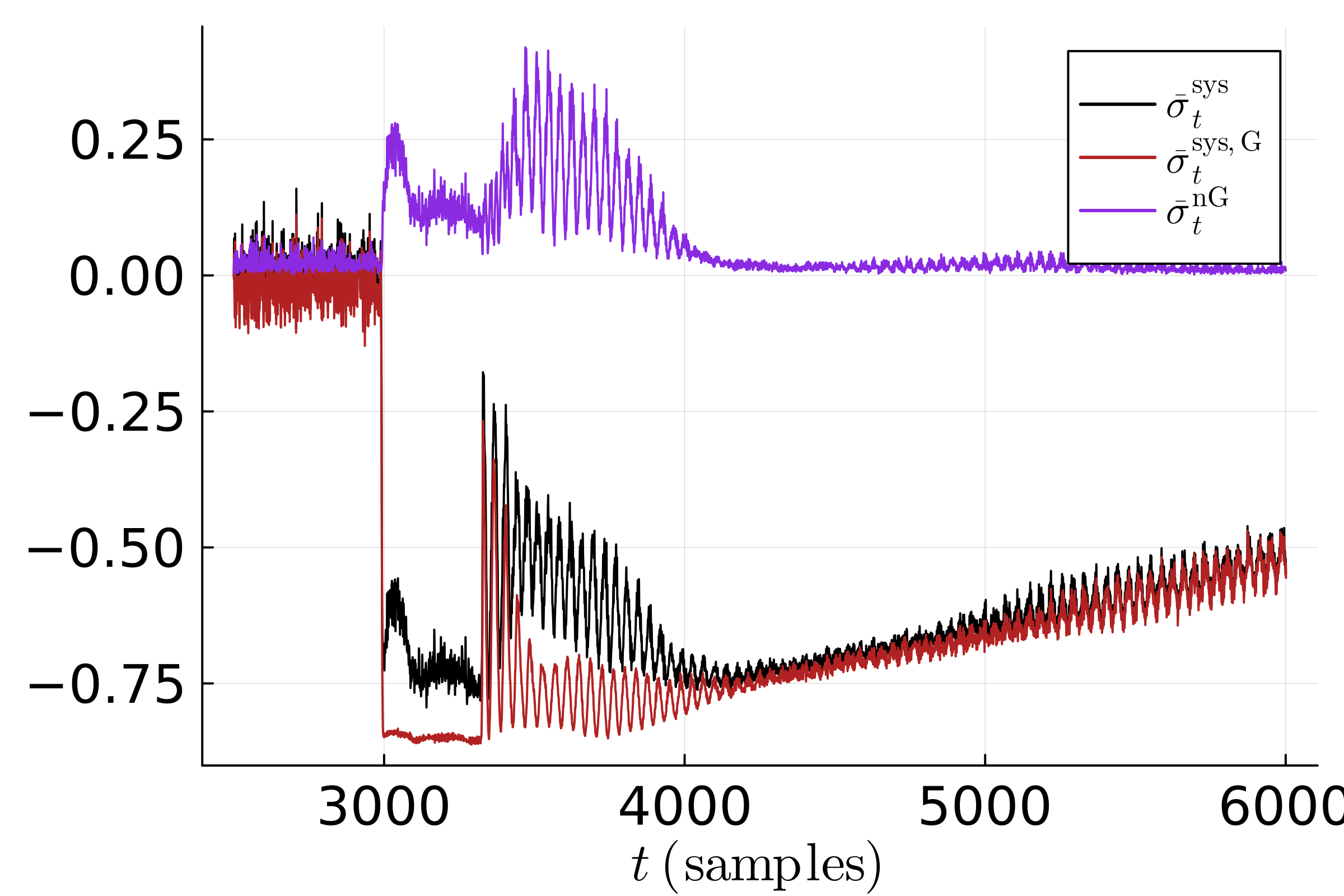}};
        \begin{scope}[x={(image.south east)},y={(image.north west)}]
            \node at (0.03,0.95) {b)};
            \draw[red,thick, ->] (0.2,0.6) -- (0.28,0.6);
            \draw[red,thick, <-] (0.36,0.6) -- (0.44,0.6);
            \node[anchor=west] at (0.45,0.6) {\Red{force pulse}};
        \end{scope}
    \end{tikzpicture}
    \begin{tikzpicture}
        \node[anchor=south west,inner sep=0] (image) at (0,0) {\includegraphics[width=\linewidth]{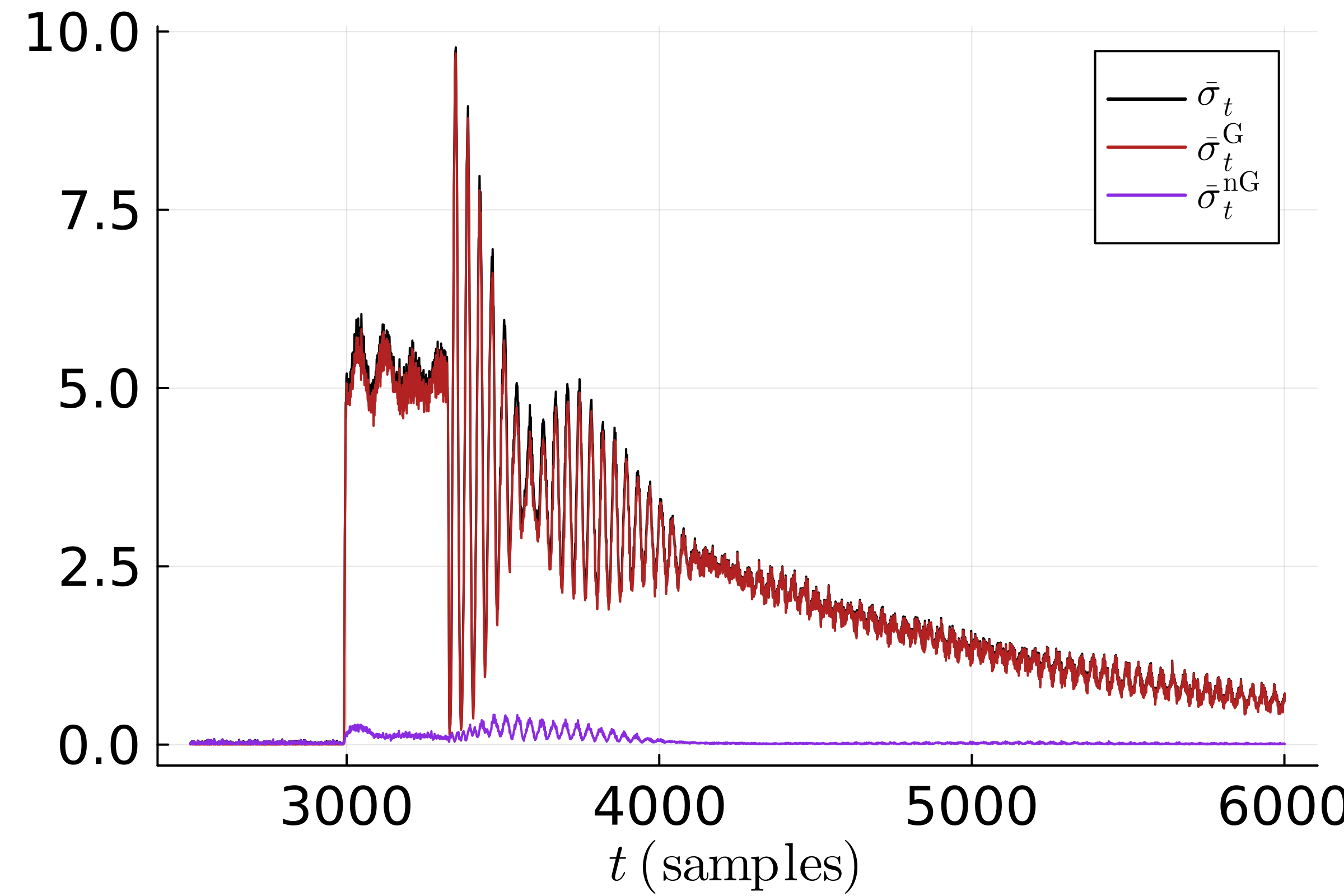}};
        \begin{scope}[x={(image.south east)},y={(image.north west)}]
            \node at (0.02,0.88) {c)};
            \draw[red,thick, ->] (0.17,0.85) -- (0.25,0.85);
            \draw[red,thick, <-] (0.33,0.85) -- (0.41,0.85);
            \node[anchor=west] at (0.42,0.85) {\Red{force pulse}};
        \end{scope}
    \end{tikzpicture}
    \caption{a) The four positive contributions to the dimensionless entropy production rate \eqref{entropy-decompositon-dimensionless} as a function of time for a trapped particle driven by an inverted-potential force pulse.
    b) The two contributions to the dimensionless rate of change of the Shannon entropy \eqref{shannon-decompositon-dimensionless} and the total value as a function of time.
    c) The overall Gaussian and non-Gaussian contribution to the entropy production rate as a function of time.
    The force pulse starts at $t = 3000$ samples and lasts $375$ samples. The sampling rate is $4.17 \cdot 10^6$ samples/s.}
    \label{fig-entropy-decop-inverted}
\end{figure}
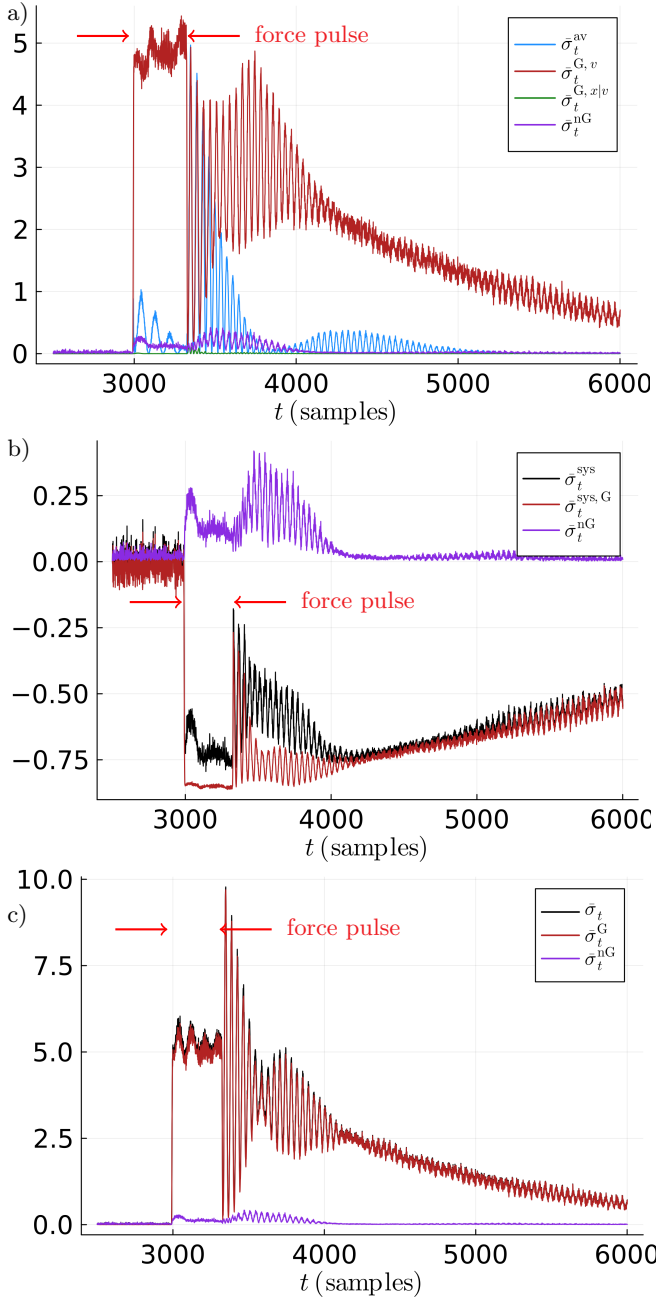

\subsection{Decomposition of entropy production in for transient, non-linear forces}

A particle that is trapped near the minimum of the optical trap will experience approximately linear forces, since the trapping potential can be approximated as parabolic \cite{Gieseler2013}, $U(x) = \kappa x^2/2$.
In equilibrium, both the position and velocity distribution are thus Gaussian,
\begin{align}
p_\text{eq}(x,v) = \frac{\sqrt{m \kappa}}{2\pi T} \exp \bigg( - \frac{m v^2}{2 T}\bigg) \exp \bigg( - \frac{\kappa x^2}{2 T} \bigg) .
\end{align}
Obviously, the entropy production rate vanishes in equilibrium.
If we were to drive the system out of equilibrium while keeping the force linear, for example, by changing the trap stiffness $\kappa$ as a function of time or applying a time-dependent bias force, then by the linearity of the Langevin equation \eqref{langevin}, the (time-dependent) probability density would remain Gaussian.
With respect to \eqref{entropy-decompositon-dimensionless}, this means that the first three terms in the decomposition can be non-zero:
A time-dependent bias force will cause the average of the particle position to depend on time, thus giving rise to a non-zero average velocity and $\bar{\sigma}_t^\text{av} > 0$.
However, a time-dependent trap stiffness will cause the width of the velocity distribution to differ from the thermal with $v_\text{th}^2$ and introduce correlations between $x$ and $v$, leading to $\bar{\sigma}_t^{\text{G},v} > 0$ and $\bar{\sigma}_t^{\text{G},x \vert v} > 0$.
However, since the joint distribution of $x$ and $v$ is Gaussian, the non-Gaussian contribution $\bar{\sigma}_t^\text{nG}$ to the entropy production rate vanishes.
The latter contribution will only appear for time-dependent and non-linear forces.
\section{Results}
As detailed in Section \ref{sec-experiment}, we drive the system out of equilibrium by exerting a non-linear force $f_t^\text{pulse}(x)$ on the particle, which, in one dimension can be written in terms of a non-parabolic potential $U_t^\text{pulse}(x)$.
We initially prepare the system in equilibrium, apply the non-linear force for a time $\tau^\text{pulse}$ and then observe the relaxation back to equilibrium.
Repeating this process allows us to obtain statistics of the particle's dynamics during the force pulse and the subsequent relaxation, which we use to compute the different contributions to entropy production rate in \eqref{entropy-decompositon-dimensionless}.
Fig.~\ref{fig-entropy-decop-quartic}a) shows the different contributions to the entropy production rate as a function of time for a force pulse with a quartic potential, $U_t^\text{pulse} = \alpha x^4/4$.
Before initiating the force pulse, all contributions to the entropy production vanish, indicating that the system is in equilibrium.
We see that the largest contribution to the entropy production rate during and after the pulse stems from the Gaussian term $\bar{\sigma}_t^{\text{G},v}$, which quantifies how much the width of the velocity distribution differs from the thermal width $v_\text{th}^2$.
Since the non-linear force strongly depends on the position of the particle, an initial, small position uncertainty leads to large differences in the force and therefore large uncertainty in the velocity.
This broadening of the velocity distribution gives the dominant contribution to dissipation.
On the other hand, the non-Gaussian contribution $\bar{\sigma}_t^\text{nG}$ to the entropy production rate is the second largest one, indicating that, indeed, the non-linearity of the force leads to a non-Gaussian velocity distribution, which in turn contributes to dissipation.

The non-Gaussian part of the entropy production rate also contributes substantially to the change in the Shannon entropy, as shown in Fig.~\ref{fig-entropy-decop-quartic}b).
Interestingly, its sign is opposite to the Gaussian part of the latter: While the broad velocity distribution due to the force pulse leads to an overall decrease in the Shannon entropy as the distribution tends to narrow during thermalization, the non-Gaussian part is always positive.
Fig.~\ref{fig-entropy-decop-quartic}b) also shows the non-monotonic behavior of the non-Gaussian contribution as a function of time during the relaxation after the force pulse.
While the non-Gaussian contribution initially decreases rapidly as the non-linear pulse force is turned off, it increases once more before slowly decaying as the particle re-thermalizes.
A possible explanation for this increase is that the trapping potential is not exactly linear:
The particle is trapped in a standing light wave, and the trapping potential is periodic in space, $U(x) = -U_0 \cos(k x)$.
Near a minimum (e.~g.~at $x = 0$), this is approximately parabolic $U(x) \simeq U_0 x^2/(2 k^2) - U_0$ and the force thus approximately linear. 
While this approximation is well justified in equilibrium, it is no longer true if the particle is displaced far from the minimum, as happens during the application of the force pulse.
As a consequence, while the particle oscillates in the trap during the relaxation, it is affected by the non-linear part of the trapping potential which also translates into a non-Gaussian velocity distribution and thus increases the non-Gaussian contribution to the entropy production rate.

When comparing the overall Gaussian contribution, that is, the first three terms in \eqref{entropy-decompositon-dimensionless}, to the non-Gaussian contribution in Fig.~\ref{fig-entropy-decop-quartic}c), we see that the Gaussian contribution dominates the overall entropy production rate.
On the one hand, this suggests that even in the presence of strongly non-linear forces, the velocity distribution does not deviate substantially from a Gaussian distribution.
On the other hand, this allows us to obtain a quantitatively accurate lower bound on the entropy production rate by considering the first three terms in \eqref{entropy-decompositon-dimensionless}.
Since these terms only depend on the first and second moments of position and velocity, they are particularly easy to evaluate from trajectory data, without the need for estimating the Fisher information.

Finally, we repeat the above analysis for a particle driven by an inverted-potential pulse $U(x) = \alpha x^4/4 - \beta x^2/2$ with $\beta > 0$.
Like a quartic potential, this potential leads to a non-linear force on the particle, with the difference that the particle is located near a maximum of the potential at the beginning of the pulse, leading to a force that pushes the particle away from its equilibrium position.
As a consequence, the overall behavior of the dissipation rate as a function of time is different.
As before, the largest contribution to the dissipation is the broadening of the velocity distribution, which enters the Gaussian part of the entropy production rate (see Fig.~\ref{fig-entropy-decop-inverted}a) and c)).
However, we now also observe a significant contribution from the average velocity of the particle.
The reason is that, since the particle is near an unstable point of the potential during the pulse, any small asymmetry of the initial distribution or misalignment of the trapping potential and the pulse potential will result in significant average motion.
Another difference to the purely quartic pulse is that in the former case the overall entropy production rate oscillates between large and small values (Fig.~\ref{fig-entropy-decop-quartic}c)) as the width of the velocity distribution oscillates due to the oscillations of the particles in the potential and transiently gets close to its equilibrium value.
By contrast, for the inverted potential, the width of the velocity distribution oscillates around a value that is larger than the equilibrium value, leading to a large positive value of the dissipation rate during the entire relaxation.
Thus, even though the rate of entropy production does not differ much between both types of pulse, the total entropy production (the integral over the entropy production rate) is larger for the inverted potential pulse, implying that the latter drives the system further from equilibrium.

We also observe an interesting behavior for the Gaussian and non-Gaussian contribution to the rate of Shannon entropy change in Fig.~\ref{fig-entropy-decop-inverted}b).
While the amplitude of the Gaussian contribution decays monotonically from a large negative value during the relaxation, the positive non-Gaussian contribution is now larger during the relaxation than during the force pulse, which we again attribute to large-amplitude oscillations which allow the particle to explore the non-linear region of the trapping potential.
As a consequence of these two competing effects, the overall rate of Shannon entropy change exhibits a strongly non-monotonic behavior during the relaxation.
As the particle returns to the linear region of the trapping potential, the non-Gaussian contribution decays quickly, even though the system remains out of equilibrium due to the broad velocity distribution.
This highlights that even for the relatively simple system of a single trapped particle, different non-equilibrium effects may contribute to dissipation in different ways and on different time scales.

\section{Discussion}
In conclusion, this study demonstrates a robust method for quantifying entropy production in weakly damped, driven systems by decomposing it into distinct contributions arising from different non-equilibrium properties of the evolution. Through the analysis of experimental trajectories of levitated nanoparticles under transient non-linear forces, we have identified that though the majority of entropy production can be accurately captured by the Gaussian contributions, specifically the non-zero mean velocity, the deviation of the width from a thermal velocity distribution, and position-velocity correlations. This suggests that the Gaussian contributions, which can be estimated from the first and second moments of position and velocity, can provide an accurate estimate on the total entropy production rate. Conversely, a notable non-Gaussian contribution was found for the change in Shannon entropy, highlighting the role of velocity statistics beyond the Gaussian approximation for a detailed picture of dissipation. Importantly, quantifying these nonlinear effects offers deeper insights into the dynamics and thermodynamics of driven systems, substantially enriching our understanding beyond standard Gaussian approximations.

From a theoretical point of view, it would be instructive to study the behavior of the different contributions to the entropy production rate in the strong damping limit.
While, to leading order, the velocity distribution becomes thermal in this limit, the entropy production in the overdamped limit originates in the sub-leading deviations from a thermal velocity distribution and thus, in principle, any of the four contributions in \eqref{entropy-decompositon-v2} may also contribute for strong damping.

Looking ahead, further experimental and theoretical developments could explore the implications of these findings in systems that involve multiple interacting particles or degrees of freedom, where nonlinear interactions and correlations can profoundly influence entropy production and non-equilibrium dynamics. While our main results also apply to these cases, such studies could shed light on the role collective phenomena and emergent behaviors arising from strong interactions and their effect on the different mechanisms of dissipation, thus broadening the applicability of our methods to diverse physical systems, from biological networks to complex engineered materials. In addition, incorporating machine-learning techniques for a more precise estimation of the Fisher information and higher-order statistical features could enhance accuracy and computational efficiency. Finally, investigating entropy production in quantum regime experiments, such as those described recently in optomechanical systems \cite{Magrini2021}, offers promising avenues for extending this approach beyond classical stochastic dynamics.

\bibliography{JabRef_main}

\appendix

\section{Estimation of velocity from time-discrete trajectories} \label{app-v-estim}
\begin{figure}
    \centering
    \includegraphics[width=0.47\textwidth]{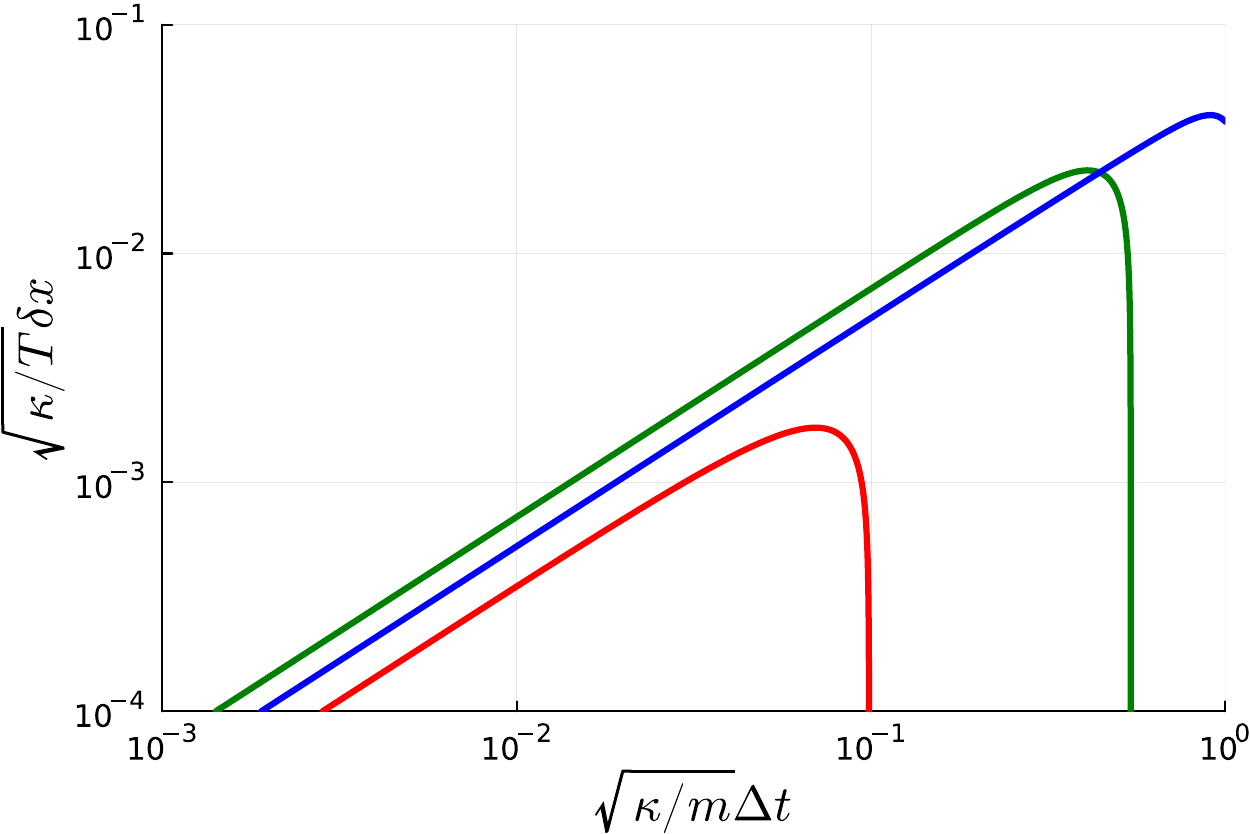}
    \caption{Estimation error of the velocity as a function of the time-resolution of the measurement $\Delta t$ (relative to the oscillation frequency in the trap) and the localization error size $\delta x$ (relative to the thermal width of the position distribution). The colored lines show the threshold for a $5\%$ error relative to the thermal velocity ($\delta v/\sqrt{T/m} < 0.05$), which is satisfied below the respective line. The red line corresponds to the forward difference \eqref{forward-difference}, the green line to the first-order central difference \eqref{central-difference} and the blue line to the second-order central difference ($k^- = k^+ = 2$).}
    \label{fig-v-error}
\end{figure}

\begin{figure*}
    \centering
    \includegraphics[width=0.47\textwidth]{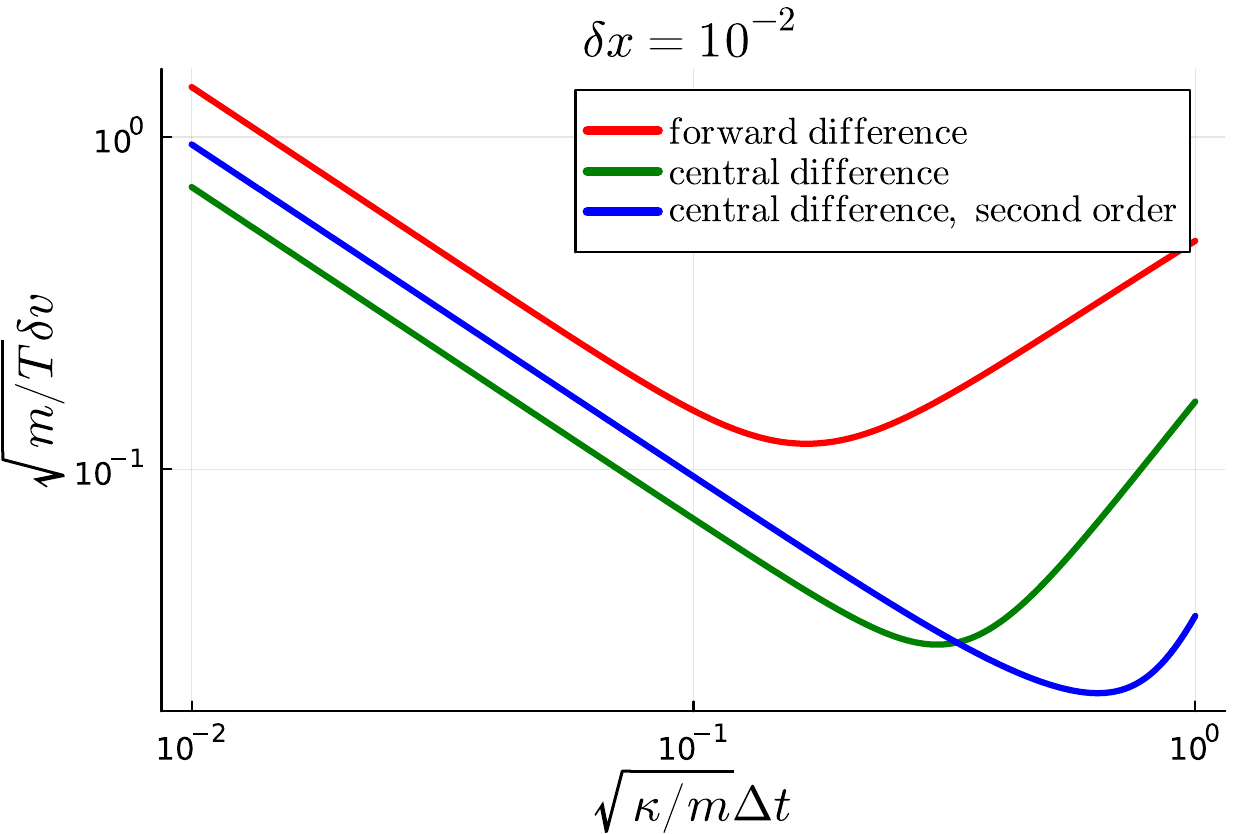}
    \includegraphics[width=0.47\textwidth]{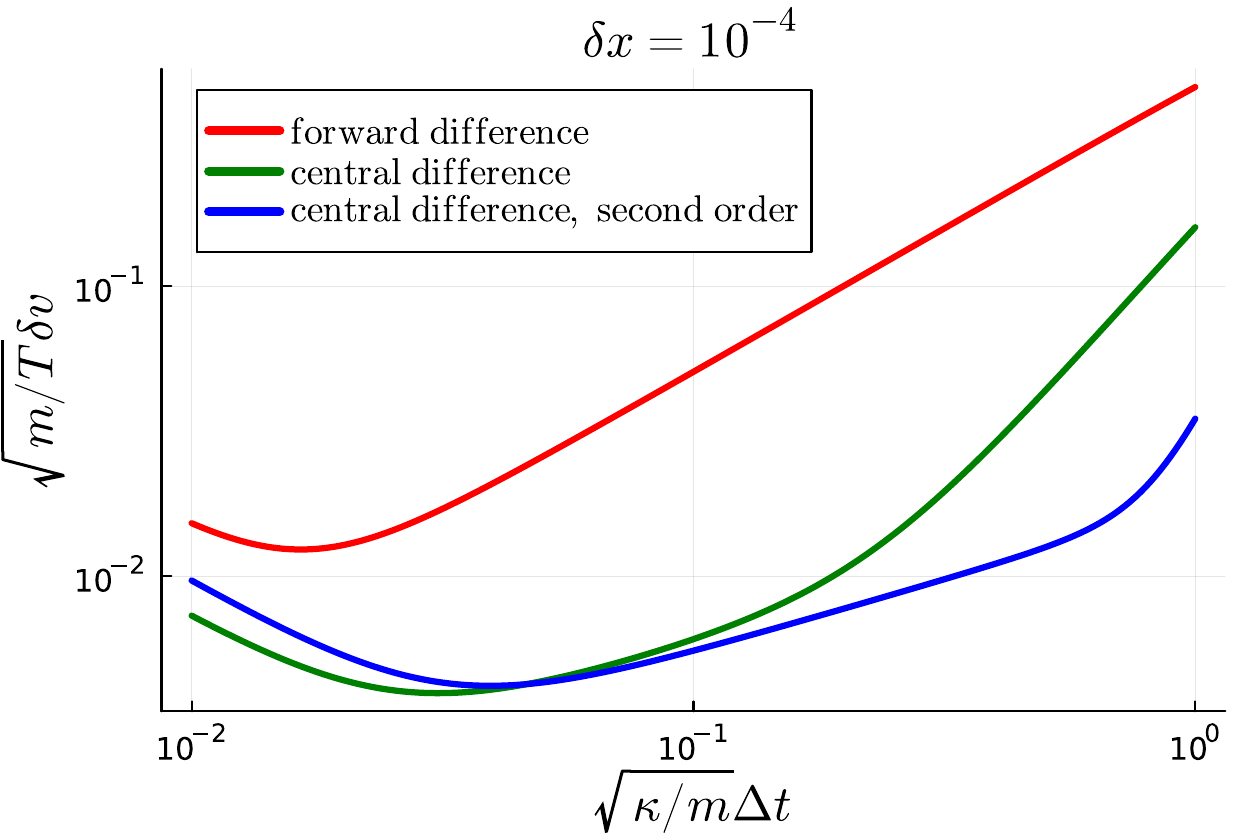}
    \caption{Estimation error of the velocity as a function of the time-resolution of the measurement $\Delta t$ (relative to the oscillation frequency in the trap) for a localization error of $\delta x = 10^{-2} \sqrt{T/\kappa}$ (left) and $\delta x = 10^{-4} \sqrt{T/\kappa}$ (right). The red line corresponds to the forward difference \eqref{forward-difference}, the green line to the first-order central difference \eqref{central-difference} and the blue line to the second-order central difference ($k^- = k^+ = 2$).}
    \label{fig-v-error-dx}
\end{figure*}

In underdamped systems, the entropy production depends crucially on the statistics of the velocity of the particle.
However, the experimental data consists of time-traces of the particle's position, from which we first have to determine the velocity.
Naively, we can estimate the velocity from subsequent measurements of the position
\begin{align}
    v(t) = \frac{x(t+\Delta t) - x(t)}{\Delta t} \label{forward-difference},
\end{align}
where $\Delta t$ is the time-resolution of the measurement.
In the absence of noise, more accurate estimates of the velocity can in principle be obtained using higher-order finite differences, for example, the central difference
\begin{align}
    v(t) = \frac{x(t+\Delta t) - x(t-\Delta t)}{2 \Delta t} \label{central-difference},
\end{align}
or more generally,
\begin{align}
    v(t) = \frac{1}{\Delta t} \sum_{k = -k^-}^{k^+} a_k x(t + k \Delta t), \label{v-estimate}
\end{align}
which uses $k^-$ past and $k^+$ future points in addition to the point at $t$, weighted by coefficients $a_k$.
The coefficients $a_k$ are the coefficients of the polynomial of order $k^- + k^+ + 1$ passing through the measurement points, also called the Lagrange polynomial.
\begin{widetext}
Concretely, for a first derivative, the coefficients can be obtained by solving the equations \cite{fdcc}
\begin{align}
    \begin{pmatrix}
        1 & 1 & 1 & \ldots & 1 & 1 \\
        (-k^-)^1 & (-k^- + 1)^1 & (-k^- + 2)^1 & \ldots & (k^+ - 1)^1 & (k^+)^1 \\
        (-k^-)^2 & (-k^- + 1)^2 & (-k^- + 2)^2 & \ldots & (k^+ - 1)^2 & (k^+)^2 \\
        \vdots & \vdots & \vdots & & \vdots & \vdots \\
        (-k^-)^{k^- + k^+} & (-k^- + 1)^{k^- + k^+} & (-k^- + 2)^{k^- + k^+} & \ldots & (k^+ -1 )^{k^- + k^+} & (k^+)^{k^- + k^+}
    \end{pmatrix}
    \begin{pmatrix}
        a_{-k^-} \\ a_{-k^- + 1} \\ a_{-k^- + 2} \\ \vdots \\ a_{k^+}
    \end{pmatrix}
    = \begin{pmatrix}
        0 \\ 1 \\ 0 \\ \vdots \\ 0
    \end{pmatrix} .
\end{align}
\end{widetext}
In practice, however, the application of higher-order methods is not always favorable due to the presence of noise.
This noise has two distinct origins: On the one hand, we have the thermal fluctuations of the particle, so that the position $x(t)$ is intrinsically noisy.
On the other hand, the measurement of $x(t)$ is not perfect, possessing a finite localization error $\delta x(t)$, so that the measured value $\hat{x}(t)$ is
\begin{align}
    \hat{x}(t) = x(t) + \delta x(t) .
\end{align}
For simplicity, we assume that $\delta x(t)$ is random, zero on average and uncorrelated with the particle position and between individual measurements, that is, $\delta x(t)$ are independent random numbers drawn from a fixed distribution.
We further assume that $x(t)$ follows the equation of motion of a damped noisy harmonic oscillator,
\begin{align}
    m \ddot{x}(t) = - \gamma \dot{x}(t) - \kappa x(t) + \sqrt{2 \gamma T} \xi(t) ,
\end{align}
and that the system is in the corresponding steady state.
We now estimate the velocity using \eqref{v-estimate} for the measured position
\begin{align}
    \hat{v}(t) = \frac{1}{\Delta t} \sum_{k = -k^-}^{k^+} a_k \hat{x}(t + k \Delta t).
\end{align}
We quantify the overall estimation error as
\begin{align*}
    \delta v = \sqrt{\Av{(\hat{v}(t) - v(t))^2}},
\end{align*}
where the average is taken with respect to the thermal noise and the localization error.
In Fig.~\ref{fig-v-error}, this error is illustrated for three finite-difference methods as a function of the time resolution and localization error.
Clearly, an accurate estimate of the velocity requires sufficient measurement time-resolution; this can be compensated to some extent by relying on higher-order finite difference methods.
However, since the localization error is independent of the measurement rate, likewise small change in actual position for small $\Delta t$ is eclipsed by the localization error.
Thus, for a given localization error, increasing the time-resolution of the measurement too much will lead to a less accurate estimate of the velocity.
Importantly, higher-order finite difference methods cannot compensate for this effect.
To the contrary, we find that the influence of the localization error is minimized for the first-order central difference \eqref{central-difference}, with higher order finite difference methods being more sensitive to localization errors.
This is shown in more detail in Fig.~\ref{fig-v-error-dx}:
For small $\Delta t$, the first order central difference has a smaller estimation error than either the forward difference or the second order central difference.
For larger localization error, the global minimum of the velocity estimation error for the second order difference is lower, however, it occurs for a time-resolution that is of the same order as the timescale of oscillations in the trap, which prohibits estimating the instantaneous velocity.
For smaller localization error, the first order difference performs better not only for small $\Delta t$ but also with respect to the minimal possible velocity estimation error.
Based on these findings, we employ the first order central difference to estimate the velocity from the trajectory data.

\end{document}